\documentclass[aps, twocolumn, nofootinbib,superscriptaddress,10pt,floatfix]{revtex4}
\usepackage{dcolumn}
\usepackage{mathrsfs}
\usepackage{graphicx}
\usepackage{amsmath,amssymb}
\usepackage{hyperref}
\usepackage{braket}
\usepackage{subfigure}
\usepackage{subfloat}
\usepackage{float}
\usepackage[dvipsnames]{xcolor}
\usepackage[normalem]{ulem}
\usepackage{placeins}

\usepackage{multirow}
\usepackage{array}

\hypersetup{
	colorlinks=true,
	linkcolor=red,
	citecolor=blue,
}

\allowdisplaybreaks

\def\be{\begin{equation}}
	\def\ee{\end{equation}}
\def\ba{\begin{eqnarray}}
	\def\ea{\end{eqnarray}}

\def\l{\left}
\def\r{\right}
\def\f{\frac}
\def\g{\gamma}

\def\etal{{\frenchspacing\it et al}}
\def\ie{{\frenchspacing\it i.e.}}

\def\mathbi#1{\textbf{\em #1}}

\begin{document}

\title{Modified recombination and the Hubble tension}

\author{Seyed Hamidreza Mirpoorian}
\email[]{smirpoor@sfu.ca}
\affiliation{Department of Physics, Simon Fraser University, Burnaby, British Columbia, V5A 1S6, Canada}

\author{Karsten Jedamzik} 
\email[]{karsten.jedamzik@umontpellier.fr}
\affiliation{Laboratoire de Univers et Particules de Montpellier, UMR5299-CNRS, Universite de Montpellier, 34095 Montpellier, France}

\author{Levon Pogosian}
\email[]{levon\_pogosian@sfu.ca}
\affiliation{Department of Physics, Simon Fraser University, Burnaby, British Columbia, V5A 1S6, Canada}

\begin{abstract}
We investigate the extent to which modifying the ionization history at cosmological recombination can relieve the Hubble tension, taking into account all relevant datasets and considering the implications for the galaxy clustering parameter $S_8$ and the matter density fraction $\Omega_m$. We use the linear response approximation to systematically search for candidate ionization histories parameterized with a cubic-spline that provide good fits to the Planck CMB and DESI BAO data while relieving the $H_0$ tension, followed by MCMC fits of the most promising candidate models to the data. We also fit to the data a physically motivated phenomenological model of ionization history that has four parameters. Our main result is that models of modified recombination can reduce the Hubble tension to below 2$\sigma$ while improving the fit to the current CMB and BAO data and reducing the $S_8$ tension. The promising candidate ionization histories have simple shapes, with no need for an oscillatory dependence on redshift. Our study also demonstrates the importance of the high-resolution CMB temperature and polarization anisotropies for constraining modified recombination, with the candidate models in this study showing varying levels of agreement with the current ACT DR4 and SPT-3G data.
\end{abstract}

\maketitle

\section{Introduction}
\label{s:intro}

The standard $\Lambda$ cold dark matter ($\Lambda$CDM) model has been very successful in providing a good fit to a wide range of cosmological observations. However, the emergence of tensions between certain datasets interpreted within the $\Lambda$CDM framework have prompted discussions of new physics. The most notable of these is the Hubble tension, which reflects a significant discrepancy between the value of the Hubble constant $H_0$ inferred from the Planck cosmic microwave background (CMB) data, $H_0 = 67.36 \pm 0.54$, and $H_0 = 73.04 \pm 1.04$ km/s/Mpc measured by the SH0ES Collaboration using Cepheid-calibrated supernovae Type Ia (SN)~\cite{Riess:2021jrx, Riess:2022mme}.

Another noteworthy discrepancy, albeit less severe, concerns measurements of the matter clustering amplitude quantified by the parameter $S_8 \equiv \sigma_8 \sqrt{\Omega_m /0.3}$, where $\sigma_8$ is the amplitude of matter density fluctuations smoothed over the scale of 8 Mpc/h, and $\Omega_m$ is the present-day matter density fraction of the critical density. In particular, the Planck-CMB-inferred value of $S_8 = 0.832 \pm 0.013$ is 1.9$\sigma$ higher than the value of $S_8 = 0.790^{+0.018}_{-0.014}$ obtained from galaxy counts and weak lensing through a joint analysis~\cite{Kilo-DegreeSurvey:2023gfr} of the Dark Energy Survey Year 3 (DES-Y3)~\cite{DES:Y3} and Kilo-Degree Survey (KiDS-1000)~\cite{KiDS:1000} data.

An additional minor tension emerged recently between the value of $\Omega_m$ deduced from uncalibrated (independent of the intrinsic supernovae brightness) SN luminosity distance vs redshift curves and that obtained from uncalibrated (independent of the comoving sound horizon at decoupling) baryon acoustic oscillations (BAO) distance measurements. The Pantheon+ (PP) SN data yield $\Omega_m = 0.334 \pm 0.018$~\cite{Brout:2022vxf}, in good agreement with DES-Y5~\cite{DES:2024jxu} and Union3~\cite{Rubin:2023ovl} SN datasets, while the recent BAO measurements by the dark energy spectroscopic instrument (DESI) give $\Omega_m = 0.295 \pm 0.015$~\cite{DESI:2024mwx}. Although only at a $\sim 2\sigma$ level, this tension adds an interesting consideration to the discussion of cosmological tensions because it is independent of the calibration of either SN or BAO.

While the possibility of unaccounted systematic effects cannot be ruled out, no single source of systematics has been identified so far that could resolve the Hubble tension \cite{Efstathiou:2020wxn, Brout:2020msh,Mortsell:2021tcx,Freedman:2021ahq} or explain the more mild $S_8$ and the $\Omega_m$ discrepancies. Naturally, the above tensions, especially the Hubble tension, have raised significant interest in potential new physics beyond the $\Lambda$CDM model. Proposed solutions to the Hubble tension are often categorized into late-time and early-time modifications~\cite{Knox:2019rjx,DiValentino:2021izs,Abdalla:2022yfr}. Late-time solutions, such as dynamical dark energy,  interacting dark matter and dark energy, modified gravity, and decaying dark matter, generally struggle to resolve the tension due to the strong constraints from the BAO and SN data~\cite{Poulin:2018zxs,Raveri:2019mxg, Benevento:2020fev,McCarthy:2022gok,Pogosian:2021mcs,Poulin:2024ken,Jiang:2024xnu,Pedrotti:2024kpn}. As a result, many researchers are focusing on the early-time solutions that reduce the comoving sound horizon $r_*$ at photon-baryon decoupling by either increasing the energy density of the Universe prior to recombination or increasing the rate of recombination. These include models of early dark energy (EDE), interacting neutrinos, varying fundamental constants, primordial magnetic fields (PMF) and others~\cite{Abdalla:2022yfr}. In particular, stochastic PMF naturally speed up recombination by inducing small-scale inhomogeneities in the baryon density prior recombination, helping relieve both the $H_0$ and $S_8$ tensions~\cite{Jedamzik:2020krr, Thiele:2021okz, Rashkovetskyi:2021rwg}.

In this paper, motivated by the PMF proposal, we investigate to what extent a modification of the recombination history can help resolve the Hubble tension without exacerbating the other tensions. Related studies were performed in~\cite{Lee:2022gzh,Lynch:2024gmp,Lynch:2024hzh}. In \cite{Lee:2022gzh}, Lee \etal \ modified recombination by allowing the electron mass $m_e$ to be a free function of redshift $z$. They used the linear response approximation (LRA) method to search for $m_e(z)$ that yield a Planck CMB best fit value of $H_0=73.04$ km/s/Mpc with the same Planck $\chi^2$ as the standard $\Lambda$CDM fit. They found that there are indeed modified recombination histories that allow for this, but not when the BAO data is added to CMB. The latter is expected, since CMB and BAO are generally consistent with each other within the standard recombination model (see \cite{Pogosian:2024ykm} for a recent analysis of this), while the two datasets require a different adjustment of $r_\star$ in order to be compatible with $H_0=73.04$ km/s/Mpc~\cite{Jedamzik:2020zmd}. Nevertheless, Lee \etal. \ found recombination histories that significantly reduce the Hubble tension while still providing acceptable fits to both CMB and BAO. In \cite{Lynch:2024gmp,Lynch:2024hzh}, Lynch \etal. \ fit a cubic spline parametrization of the ionized fraction, $x_e(z)$, to the combination of Planck and DESI BAO (without a SH0ES prior on $H_0$) and found that this yields $H_0 = 70 \pm 1.1$ km/s/Mpc, providing a considerable reduction of the Hubble tension.

Our study builds on the approaches of \cite{Lee:2022gzh,Lynch:2024gmp,Lynch:2024hzh} in two ways. First, we revisit the LRA method used in \cite{Lee:2022gzh} and, instead of $m_e$, vary $x_e(z)$ directly to systematically search for parameter combinations that yield $H_0>70$ km/s/Mpc, while improving the $\chi^2$ and reducing the value of $S_8$. We find that one can have $H_0 \sim 71$  km/s/Mpc with a lower CMB+BAO $\chi^2$ than the $\Lambda$CDM best fit. The LRA method, as well as the cubic spline parametrization of \cite{Lynch:2024gmp,Lynch:2024hzh}, yield wiggly $x_e(z)$ that would be difficult to realize in a physical system. While Lynch {\it et al}~\cite{Lynch:2024hzh} found that smoothing over the wiggles of their best-fit $x_e(z)$ only marginally degraded the quality of the fit, it is interesting to understand to what extent having the wiggles is helpful for achieving better fits with higher values of $H_0$. To address this question, we introduce a simple four-parameter model of $x_e(z)$ that can approximately match the recombination histories predicted by PMF and fit it to the data. Without the wiggles in $x_e(z)$, we find that one can in fact improve the CMB+BAO fit, while getting $H_0 \sim 71$ km/s/Mpc and reducing the $S_8$ tension. We also demonstrate the important role of small-scale CMB temperature and polarization anisotropy spectra in discriminating between different modified recombination models.

In what follows, Sec.~\ref{s:methodology} introduces our LRA-based parameter search and the 4-parameter model, Sec.~\ref{s:Data} describes the datasets used in the analysis, followed by the results presented in Sec.~\ref{s:results} and a discussion in Sec.~\ref{s:discussion}. We conclude with a brief summary in Sec.~\ref{s:summary}. Additional analysis details and results are presented in the Appendixes. 

\section{Methodology}
\label{s:methodology}

The more successful theoretical proposals to resolve the Hubble tension involve reducing the comoving sound horizon at the photon-baryon decoupling, $r_\star$. This shifts the acoustic peaks in the CMB and matter power spectra to smaller scales, requiring a reduction in the comoving distance to the redshift of decoupling, and hence a larger $H_0$, in order to preserve the observed angular scales of the acoustic features. The sound horizon is computed as
\be
\label{eq:rstar}
r_\star = \int_{z_\star}^{\infty} \f{c_s(z)}{H(z)} \,dz,
\ee
where $c_s(z)$ is the sound speed of the baryon-photon fluid, $H(z)$ is the redshift-dependent cosmological expansion rate, and $z_\star$ is the redshift of decoupling that depends on the recombination history. The latter is described by the ionized fraction $x_e(z)$, defined as
\be
\label{eq:xe}
x_e(z) = \f{n_e}{n_H},
\ee
where $n_e$ is the number density of free electrons, and $n_H$ is the total number density of hydrogen nuclei in the primordial plasma. The ionization fraction determines the differential optical depth $\dot{\tau}$, given by $\dot{\tau} = \sigma_T n_e a$, where $\sigma_T$ is the Thomson scattering cross section, and $a$ is the scale factor. The visibility function, which describes the probability of last photon scattering at a given redshift, is given by $g(z) \equiv \dot{\tau} e^{-\tau}$. The peak of the visibility function defines the redshift of photon decoupling, $z_\star$, which sets the bound in the integral (\ref{eq:rstar}) and marks the time when photons last scattered and the CMB was released~\cite{Hu:1995en}. Thus, a change in $x_e(z)$ would affect $z_\star$ and, therefore, $r_\star$ and the inferred value of $H_0$. 

We adopt two distinct yet complementary approaches to modifying the ionization history in order to address cosmological tensions, with a particular focus on the Hubble tension. First, we use the LRA method to search in a systematic way for small changes in the ionization fraction $x_e(z)$ across a broad redshift range that help relieve the $H_0$ and $S_8$ tensions while maintaining a good fit to the CMB and BAO data. Then we consider a four-parameter model of $x_e(z)$ motivated by baryon clumping models in the studies of PMF. Both approaches are designed to assess the extent to which modifications to the recombination history can relieve the Hubble tension in a model-agnostic way.

\subsection{The linear response approximation method}
\label{ss:LRA_model}

One way to search for modified recombination histories that solve the Hubble tension is by adopting a flexible parametrization of $x_e(z)$, {\it e.g.}, a cubic spline with many nodes, and fitting it to data to explore the acceptable parameter space. In practice, this is computationally expensive, and can be plagued by degeneracies due to the large number of correlated parameters. One can address the computational cost through the use of emulators, as was done in \cite{Lynch:2024gmp,Lynch:2024hzh}. Still, this procedure would only provide a best fit and marginalized posterior distributions for the $x_e(z)$ parameters, potentially missing viable $x_e(z)$ shapes that give acceptable fits but not necessarily the best fit. 

The LRA method offers a quick way to search in a systematic way for all acceptable $x_e(z)$. It employs the Fisher approximation to compute the changes in the minimum $\chi^2$ due to small changes in parameters around a given fiducial model. The parameters could include nodes of a smooth function, such as $x_e(z)$, making it possible to predict how a change in $x_e(z)$ would affect the $\chi^2$ without the computational cost of performing a MCMC fit to the data. The limitation of the LRA method is that the accuracy of the prediction is reduced when departures from the fiducial model are large. However, one can always test the quality of the LRA prediction by fitting a few predicted histories that reduce the cosmological tensions while improve the $\chi^2$ to the data, which is the approach we take in this study.

A general observable can be represented by a vector $\mathbi{X}$, which may consist of multiple observables such as the CMB temperature and polarization power spectra, BAO measurements, {\it etc}. In our analysis, $\mathbi{X}$ includes the Planck CMB temperature and polarization spectra, and the BAO measurements from DESI. The theoretical prediction $\mathbi{X}^{\rm{th}}$ depends on a set of standard cosmological parameters given by a vector $\vec{\Omega} \equiv \l\{ \Omega_c h^2, \Omega_b h^2, H_0, \tau, \mathrm{log}(10^{10} A_s), n_s \r\}$. Assuming a Gaussian likelihood, the $\chi^2$ for a theoretical model can be written as
\be
\label{eq:chi_squared}
\chi^2 (\vec{\Omega}) = \l[ \mathbi{X}^{\rm{th}} (\vec{\Omega})
- \mathbi{X}^{\rm{obs}} \r] . \, \mathbi{M}(\vec{\Omega}) \, . \l[ \mathbi{X}^{\rm{th}} (\vec{\Omega})
- \mathbi{X}^{\rm{obs}} \r],
\ee
where $\mathbi{X}^{\rm{obs}}$ is the observed data vector, and $\mathbi{M} = \mathbf{\Sigma}^{-1}$ is the inverse of the covariance matrix. In \cite{Lee:2022gzh}, Lee \etal. have shown that the shifts in the best-fit parameters and $\chi^2$ due to changes in a smooth function can be calculated using integrals involving functional derivatives with respect to that function. Replacing the electron mass $m_e(z)$ with $\mathrm{ln} \, [x_e(z)]$, and otherwise closely following their formalism, the change in the $i$th best-fit parameter $\Omega^{i}_{\mathrm{BF}}$ due to a perturbation of the ionization history $\Delta \mathrm{ln} \, [x_e(z)]$ can be written as
\be
\label{eq:delta_omegabf}
\Delta \Omega^{i}_{\mathrm{BF}} = -\int dz~ F^{-1}_{ij} \f{\partial \mathbi{X}}{\partial \Omega^j} \, . \, \mathbi{M} \, . \,
\f{\delta \mathbi{X}}{\delta \mathrm{ln} \, [x_e(z)]} ~\Delta \mathrm{ln} \, [x_e(z)]
\ee
where all the quantities inside the integral except $\Delta \mathrm{ln} \, [x_e(z)]$ are evaluated at the fiducial cosmology that is assumed to be close to the new best fit. We take the fiducial model as the best fit $\Lambda$CDM model with the conventional recombination history. Here $F_{ij}$ is the Fisher matrix evaluated at the fiducial model given by
\be
\label{eq:fisher}
F_{ij} = \l( \f{\partial \mathbi{X}}{\partial \Omega^i} \, . \, \mathbi{M}(\vec{\Omega}) \,
. \, \f{\partial \mathbi{X}}{\partial \Omega^j} \r) \bigg|_{\rm{fid}}.
\ee
One can also derive an expression for the change in the best-fit chi-squared, $\Delta \chi^2_{\rm BF}$, due to the perturbation $\Delta \mathrm{ln} \, [x_e(z)]$ that we provide in Appendix \ref{ss:shifts}. The optimization problem then reduces to finding $\Delta \mathrm{ln} \, [x_e(z)]$ such that $\Delta \chi^2_{\rm BF} \le 0$, while providing a relief of the cosmological tensions. In particular, we will be searching for all $\Delta \mathrm{ln} \, [x_e(z)]$ that reduce $\chi^2_{\rm BF} $ while predicting $H_0^{\rm BF} \ge 70$, and $S_8^{\rm BF} < S_{8}^{\rm{fid}}$.

To evaluate the integral in Eq.~(\ref{eq:delta_omegabf}) and the corresponding integral for $\Delta \chi^2_{\rm BF}$, first we need to compute the functional derivatives such as ${\delta \mathbi{X}}/{\delta \mathrm{ln} \, [x_e(z)]}$. We evaluate them numerically using the formalism used by Hart \etal. in \cite{Hart:2019gvj}. Namely, we create an orthonormal set of basis functions of Gaussian shape centered at a set of redshift values $z_k$ to perturb $x_e(z)$ in narrow redshift ranges around $z_k$. We then implement the perturbed $x_e(z)$ in the recombination routine {\tt RECFAST}~\cite{Seager:1999bc,Wong:2007ym,2011ascl.soft06026S} of the Boltzmann code {\tt CAMB}~\cite{2000ApJ...538..473L} and compute the response of observables $\mathbi{X}$ to the perturbation. Having evaluated all necessary functional derivatives, we can compute $\Delta \Omega^{i}_{\mathrm{BF}}$ and $\Delta \chi^2_{\rm BF}$ for any (small) modification of ionization history $\Delta \mathrm{ln} \, [x_e(z)]$. With this pipeline at hand, we can search for $\Delta \mathrm{ln} \, [x_e(z)]$ that satisfy our criteria. We do this by generating a large number of $\Delta \mathrm{ln} \, [x_e(z)]$ histories using cubic splines with 7 control points placed evenly between $z = 700$ and $z = 1600$. Further details of our procedure are given in Appendix \ref{app:numerics}.

The accuracy of the LRA method worsens for best fits that are further away from the fiducial model. To check how well the LRA-predicted best fit ionization histories actually fit the data, we perform MCMC fits in a few cases that provide the most promising results. In particular, for each data combination,  we select one LRA-predicted history that yields a reasonable, though not unique, combination of high $H_0$, low $S_8$, and low $\chi^2$, prioritizing cases with high $H_0$ and low $S_8$ values. When testing this LRA prediction with MCMC, we implement the corresponding $\Delta \ln x_e(z)$ as the modified recombination model in RECFAST. In addition to checking how well a given LRA prediction actually fits the combination of Planck CMB and DESI BAO data, we also check the impact of adding the SN data and the high-resolution CMB anisotropy measurements to the dataset.

\subsection{The four-parameter model}
\label{ss:4model}

Fully agnostic parametrizations of $x_e(z)$, such as a cubic spline, are helpful for exploring the extent to which modified recombination can relieve the tensions. However, such methods tend to yield best fit ionization histories that would be difficult to realize in a physical system, such as the highly oscillatory $x_e(z)$ obtained in \cite{Lee:2022gzh,Lynch:2024gmp,Lynch:2024hzh} and our LRA results presented in Sec.~\ref{s:results}. It is, therefore, interesting to know if modifications of simpler shape could help relieve the Hubble tensions at a similar level.

To this end, we adopt a four-parameter model of $x_e(z)$ motivated by ionization histories obtained from simulations of recombination in the presence of PMF~\cite{Jedamzik:2023rfd}. The two key features of a generic PMF-induced relative change in $x_e(z)$ is an overall shift of recombination to higher redshifts along with a bump $1000 <z<1200$. The four-parameter model allows for an overall shift of the unmodified ionization history $x_e^{(0)}$ in redshift quantified by $\Delta z_{\mathrm{shift}}$, along with a Gaussian-shape bump at a certain reference redshift $z_{\mathrm{b}}$ of amplitude $A_{\mathrm{b}}$ and a width-at-half-maximum $\sigma_{\mathrm b}$. It is given by
\be
\label{eq:xe4model}
x_e(z) = x_e^{(0)}(z - \Delta z_{\mathrm{shift}}) \l\{ 1 + A_{\mathrm{b}} \, \mathrm{exp} \l[ - \f{(z - z_{\mathrm{b}})^2}{2 \sigma_{\mathrm b}^2} \r] \r\},
\ee  
where, for a given set of cosmological parameters, $x_e^{(0)}$ is the output of the standard recombination model implemented in {\tt RECFAST}.

\section{Datasets}
\label{s:Data}

In this section we separately describe the datasets in the data vector and the Fisher matrix used in the LRA forecast, and the datasets used in the MCMC analysis when fitting the LRA predictions and the four-parameter to data.

\subsection{The data vector in the LRA method}
\label{ss:lra-data}

The data vector $\mathbi{X}^{\rm obs}$ in our LRA Fisher forecast consists of the Planck CMB temperature and polarization spectra, and the BAO measurements from DESI. We use this combination because the extent to which on can relieve the Hubble tension by reducing the sound horizon is primarily limited by the challenge of maintaining simultaneously good fits to the CMB and BAO data. This will be further discussed in Sec. \ref{s:discussion}.

\subsubsection{Planck CMB}
\label{sss:planck}

We use both the high-$\ell$ and low-$\ell$ Planck CMB likelihoods in our LRA Fisher forecast. For $\ell \ge 30$, we use the $C^{TT}_{\ell}$, $C^{TE}_{\ell}$, and  $C^{EE}_{\ell}$ spectra and the covariance matrix from the {\tt Cobaya} \cite{Torrado:2020dgo, 2019ascl.soft10019T} implementation of the Planck PR4 CamSpec ({\tt NPIPE}) likelihood \cite{Rosenberg:2022sdy}. The high-$\ell$ CMB part of the data vector in the LRA analysis is given by
\be
\label{eq:x_high}
\mathbi{X}^{\rm obs}_{\mathrm{high}-{\ell}} \equiv  \l\{ C^{TT}_{\ell(143)}, C^{TT}_{\ell(217)}, C^{TT}_{\ell(143 \times 217)}, C^{TE}_{\ell}, C^{EE}_{\ell}\r\} \ ,
\ee
where we used $C^{TT}_{\ell}$ from two high-frequency maps at $143$ and $217$ GHz, and the $143 \times 217$ cross spectra, with $\ell_{\rm max} = \{2000, 2500, 2500\}$, respectively. To compute the difference between $\mathbi{X}^{\rm obs}$ in (\ref{eq:x_high}) and ${\mathbi{X}^{\rm th}}$ in (\ref{eq:chi_squared}), one needs to add the contribution from foreground residuals to the fiducial cosmology prediction. We use the expression given by \cite{Rosenberg:2022sdy}
\be
\label{eq:foregrounds}
C_\ell^{\rm power} = A_{\nu}^{\rm power} \l( \f{\ell}{1500} \r)^{\g_{\nu}^{\rm power}},
\ee 
where the nuisance parameters $A_{\nu}^{\rm power}$ and $\g_{\nu}^{\rm power}$ represent the amplitudes and spectral indices of power-law fits to foreground residuals for each of the TT spectra labeled by $\nu$.

For the temperature spectra $C^{TT}_{\ell}$ with $\ell < 30$ we use the {\tt Cobaya} implementation of the Planck 2018 baseline low-$\ell$ TT likelihood~\cite{Planck:2018vyg}.

For the $\ell < 30$ E-mode polarization power spectra, we use {\tt Planck-low-py}, a Python-based compressed likelihood introduced by Prince et al.~\cite{Prince:2021fdv}, which has been shown to provide constraints on cosmological parameters consistent with the full likelihoods from the Planck legacy release. In their formalism, the EE spectra are compressed into three bins, and a log-normal distribution is used to approximate the likelihood. The probability distribution for the binned power spectrum $D_{\ell}^{EE} = \ell (\ell + 1) C_{\ell}^{EE}/2\pi$ is given by
\be
\label{eq:L_ee}
\mathcal{L}(x) = \f{1}{(x - x_0) \sigma \sqrt{2\pi}} \mathrm{exp} \l[ -\f{(\mathrm{ln}(x - x_0) - \mu)^2}{2\sigma^2} \r],
\ee
where $x=D^{EE}_{\rm bin}$ with the best-fitting parameters $x_0$, $\mu$, and $\sigma$ provided in \cite{Prince:2021fdv}. We use $\mathbi{X}^{EE}_{\mathrm{low}-{\ell}} \equiv D^{EE}_{\rm bin}$ as the low-$\ell$ EE contribution to the data vector in the LRA analysis.

\subsubsection{DESI 2024 BAO}
\label{sss:desi_bao}

We use the DESI Year 1 BAO data~\cite{DESI:2024mwx} comprised of three types of observables: $D_{\rm M}(z)/r_{\rm drag}$, $D_H(z)/r_{\rm drag}$, and $D_{\rm V}(z)/r_{\rm drag}$. Here $r_{\rm drag}$ is the sound horizon at the ``baryon drag'' epoch, with $r_{\rm drag} \approx 1.02 r_\star$ in $\Lambda$CDM, $D_{\rm M}$ is the comoving distance to redshift $z$, $D_H = c/H(z)$, and $D_{\rm V}(z) = [z D^2_{\rm M}(z) D_H (z)]^{1/3}$~\cite{SDSS:2005xqv}. The BAO contribution the data vector consists of 12 data points of these three types at 7 effective redshifts, $z_{\rm eff}$= 0.295, 0.510, 0.706, 0.930, 1.317, 1.491, and 2.330, taken from \cite{DESI:2024mwx}.

\subsection{Datasets used in MCMC runs}
\label{ss:mcmc_datasets}

In our MCMC fits of the four-parameter model and the ionization histories predicted by the LRA method to the data we use combinations of the following datasets:
\begin{itemize}
	\item \textbf{Planck}:  the Planck PR4 CamSpec ({\tt NPIPE}) likelihood \cite{Rosenberg:2022sdy} for the $\ell \ge 30$ CMB TT, TE and EE spectra, the Planck 2018 baseline $\ell<30$ TT and EE likelihoods~\cite{Planck:2018vyg}, and the Planck PR4 lensing likelihood~\cite{Carron:2022eyg};
	\item \textbf{DESI}: the 12 BAO measurements at 7 effective redshifts from DESI Year 1 release~\cite{DESI:2024mwx};
	\item \textbf{SPT}: the high-resolution CMB TT, TE, EE spectra from the 2022 release of the South Pole Telescope 3G (SPT-3G) experiment~\cite{SPT-3G:2022hvq};
	\item \textbf{ACT}: high-resolution CMB TT, TE, EE spectra from the 4th data release of the Atacama Cosmology Telescope experiment (ACT-DR4)~\cite{ACT:2020gnv};
	\item \textbf{PP}: The Pantheon+ collection of uncalibrated SN magnitudes~\cite{Brout:2022vxf};
	\item \textbf{PP}{\boldmath$M_b$}: The Pantheon+ (PP) SN dataset~\cite{Brout:2022vxf} calibrated using the brightness magnitude $M_b = -19.253 \pm 0.027$ from the cosmic distance ladder measurement by the SH0ES collaboration~\cite{Riess:2021jrx}.
\end{itemize}
We use {\tt Cobaya}~\cite{Torrado:2020dgo, 2019ascl.soft10019T} to perform the MCMC analysis.

\section{Results}
\label{s:results}

\begin{figure*}[!htbp]
	\centering
	\begin{subfigure}
		\centering
		\includegraphics[width=0.9\textwidth]{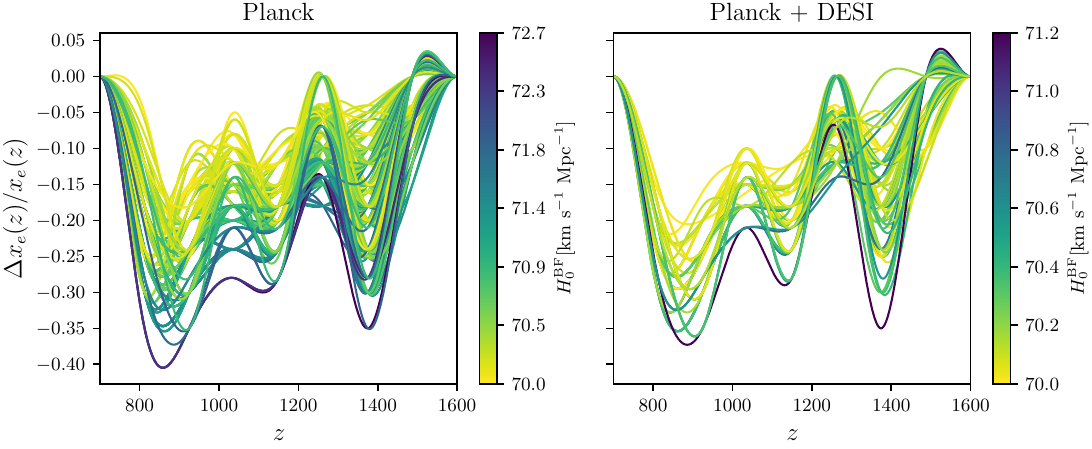}
		\label{fig:lnx_H0}
	\end{subfigure}
	\begin{subfigure}
		\centering
		\includegraphics[width=0.9\textwidth]{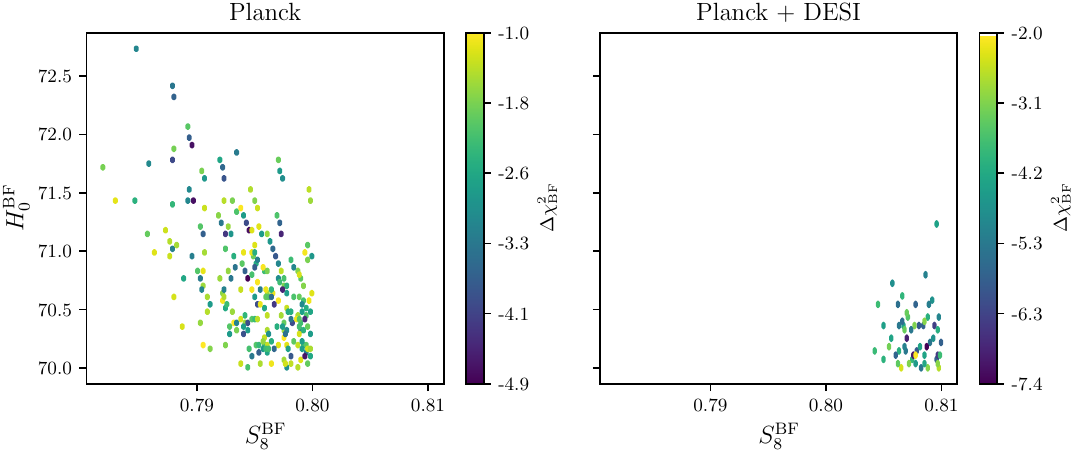}
		\label{fig:H0_s8}
	\end{subfigure}
	\caption{Top: LRA-predicted ionization histories, plotted in terms of their relative difference from $\Lambda$CDM, with the best-fit $H_0$ values indicated with the color bar, for Planck (left) and Planck + DESI (right). Bottom: the values of $H_0$ and $S_8$ for the ionization histories shown in the top panels with the color bar indicating the corresponding $\Delta \chi^2_{\rm BF}$ values. We restricted our search to models with $S_8^{\rm BF} < 0.8$ for Planck, and $S_8^{\rm BF} < 0.81$ for Planck + DESI.}
	\label{fig:lnx_lra}
\end{figure*}

\begin{figure*}[!htbp]
	\centering
	\includegraphics[width=0.9\textwidth]{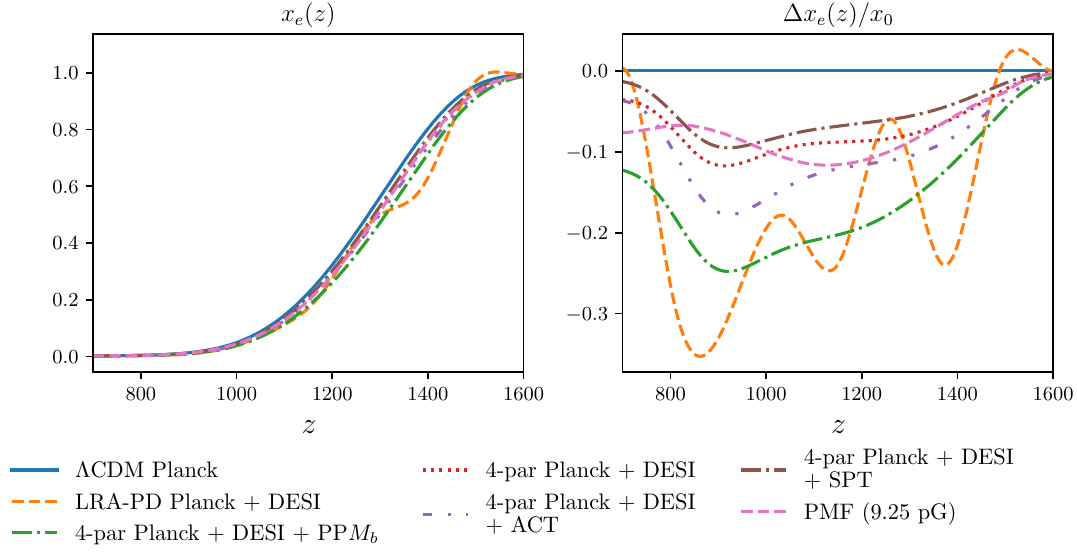}
	\caption{The ionization histories $x_e(z)$ predicted by the LRA based on the best fit to Planck+DESI (orange dash) and the four-parameter model best fits to Planck+DESI (red dot), Planck+DESI+PP$M_b$ (green dash-dot), Planck+DESI+ACT (purple dash-double-dot) and Planck+DESI+SPT (brown dash-dot). The Planck best fit $\Lambda$CDM ionization history is shown with a solid blue line. For comparison, the pink dashed line shows the $x_e(z)$ predicted by PMF of $B_{\rm rms} = 9.25\,$pG that gives a good fit to the combination of Planck and DESI.}
	\label{fig:recom_fits}
\end{figure*}

\begin{figure*}[!htbp]
	\centering
	\begin{subfigure}
		\centering
		\includegraphics[width=0.48\textwidth]{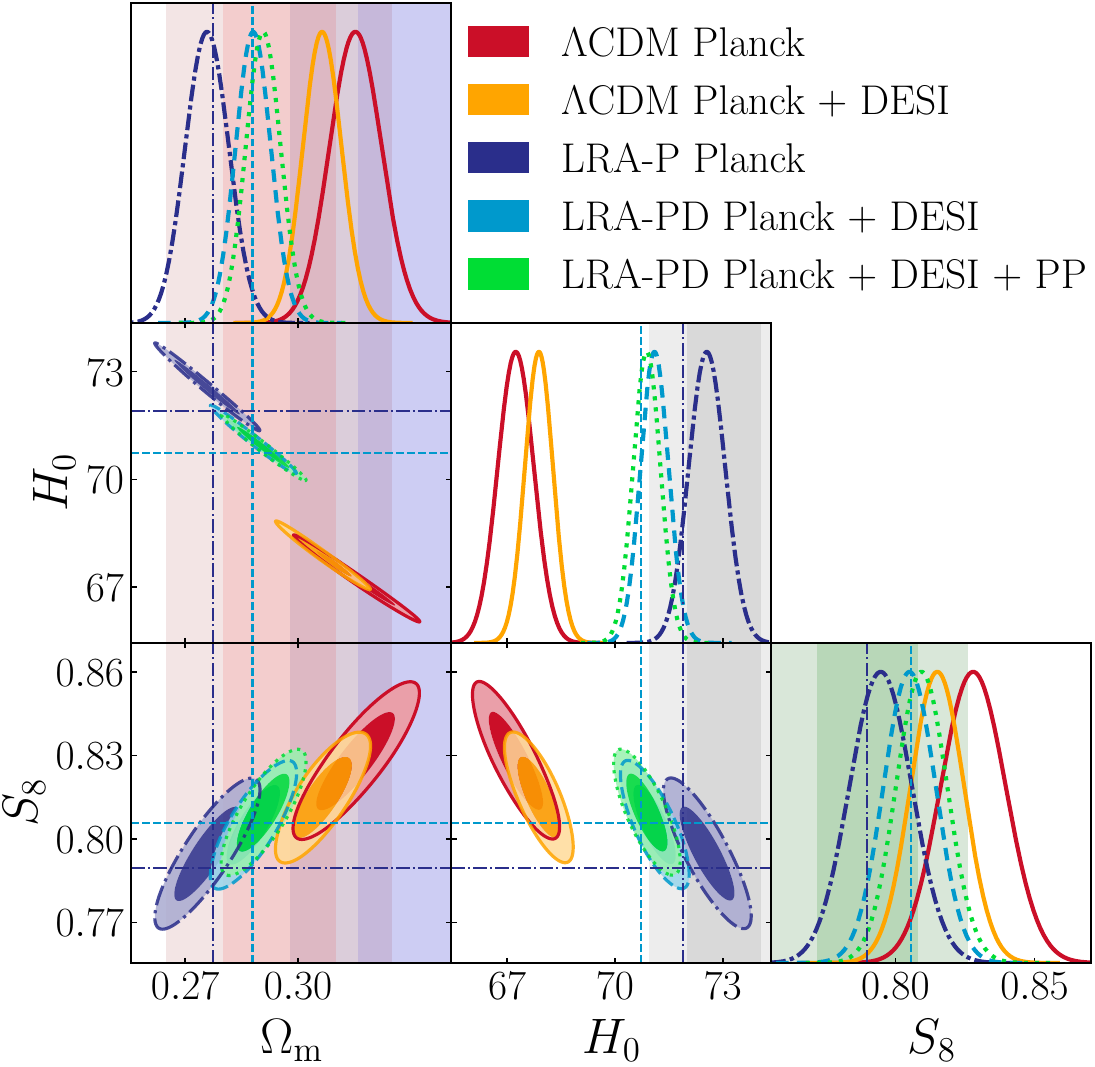}
	\end{subfigure}
	\hspace{0.1cm}
	\begin{subfigure}
		\centering
		\includegraphics[width=0.48\textwidth]{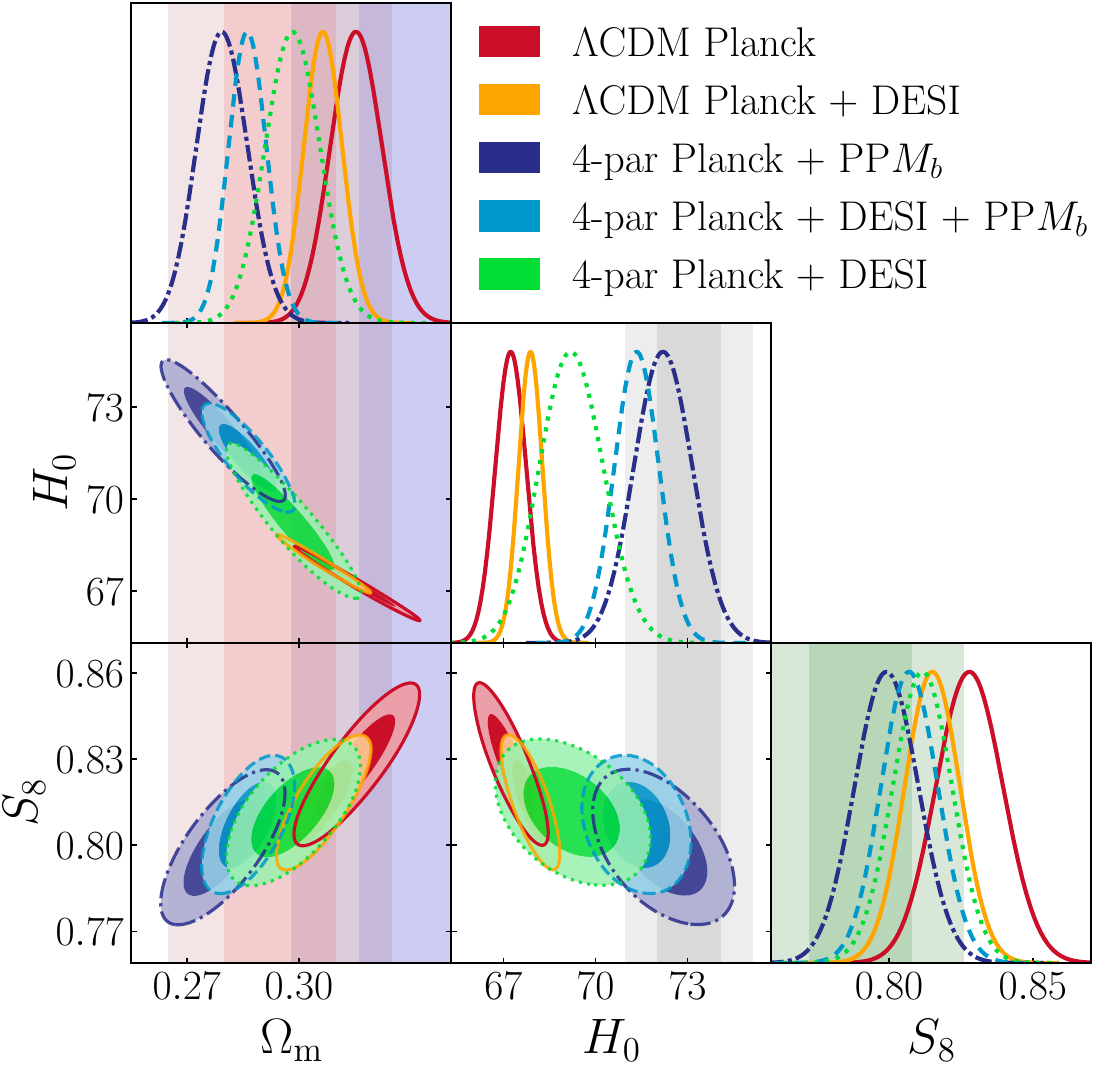}
	\end{subfigure}
	\caption{The 68\% and 95\% CL contours of the $\Omega_m$, $H_0$, and $S_8$ posterior distributions for the LRA-P and LRA-PD models (left) and the four-parameter model (right) fit to combinations of Planck, DESI, PP and PP$M_b$ as discussed in the text. For reference, the red, blue, gray and green vertical bands show the 68\% and 95\% CL intervals for $\Omega_m = 0.295 \pm 0.015$ from DESI, $\Omega_m = 0.334 \pm 0.018$ from PP, $H_0 = 73.04 \pm 1.04$ from SH0ES, and $S_8 = 0.790^{+0.018}_{-0.014}$ from DES-Y3 + KIDS-1000 respectively. The dark(light) blue dash-dotted (dashed) lines show the best fit values predicted by the LRA method for Planck (Planck+DESI).}
	\label{fig:contours_all}
\end{figure*}

As described in Sec.~\ref{ss:LRA_model}, we use the LRA method to search for $x_e(z)$ that reduce the best-fit Planck and DESI chi-squared relative to the fiducial $\Lambda$CDM model, while predicting $H_0^{\rm BF} \ge 70$ and $S_8^{\rm BF} < S_{8}^{\rm{fid}}$. We perform this search separately for Planck alone and for the combination of Planck and DESI by generating a large number of ionization histories using cubic splines with 7 control points placed evenly between $z = 700$ and $z = 1600$. 

The top panels in Fig.~\ref{fig:lnx_lra} show the recombination histories obtained via the LRA method that satisfy our three criteria with the colors indicating the corresponding values of $H_0$. The bottom panels show the distribution of the $H_0$ and $S_8$ values for the same histories, with the colors indicating the improvement in $\chi^2_{\rm BF}$. It is worth noting that the success rate of finding recombination histories which pass all three criteria was very small, $\sim 0.02\%$ for Planck and $\sim 0.005\%$ Planck + DESI. This suggests that relieving the Hubble tension requires a very particular modification of the recombination history. It is also apparent that the inclusion of the BAO data yields smaller $H_0$ and larger $S_8$, as expected for reasons presented in~\cite{Jedamzik:2020zmd}.

The successful recombination histories in Fig.~\ref{fig:lnx_lra} are generally wiggly. We attribute this, in part, to our choice of the parametrization, {\it i.e.} using a cubic spline with seven pivot points. As we discuss later in this section, the presence of the wiggles is not a necessary condition for modified recombination to substantially relieve the $H_0$ tension.

Next, we select two specific ionization histories obtained from the LRA search: one from the analysis of the best fit to Planck that predicts $H_0 = 71.9$ km/s/Mpc, $S_8 = 0.79$, and $\Delta \chi^2_{\rm BF} = -4.71$, and one from the combination of Planck and DESI that predicts $H_0 = 70.74$ km/s/Mpc, $S_8 = 0.806$, and $\Delta \chi^2_{\rm BF} = -4.77$. We will treat these two ionization histories as two specific models of modified recombination, labeled as ``LRA-P'' and ``LRA-PD'', that we will fit to the data alongside the $\Lambda$CDM and the four-parameter model. We do this in order to test the accuracy of the LRA prediction, examine the importance of the wiggles in $x_e(z)$, and investigate the impact of adding other datasets, such as ACT, SPT, PP and PP$M_b$. The LRA-PD $x_e(z)$ is shown in Fig.~\ref{fig:recom_fits}.

Figure~\ref{fig:contours_all} shows the marginalized posteriors for $\Omega_{\rm m}$, $S_8$, and $H_0$ obtained from the MCMC fits of the LRA-P and LRA-PD models, and the four-parameter model, to several data combinations. The mean values and 68\% confidence level (CL) uncertainties in these and other relevant parameters, along with the Planck and DESI BAO best fit $\chi^2$ are provided in Table~\ref{t:parameters_all}. In the case of the two LRA models, the LRA predicted values of these three parameters are shown with dash-dotted (dashed) vertical and horizontal lines in the left panel of Fig.~\ref{fig:contours_all}. The posteriors of the LRA-P model parameters fit to Planck show that the reduction in the $H_0$ and $S_8$ tensions is not as good as predicted by the LRA, but still considerable. We find $H_0 = 72.57 \pm 0.47$ km/s/Mpc, while reducing the Planck $\chi^2$ by $2.92$ relative to the $\Lambda$CDM fit to Planck. This model also predicts $\Omega_{\rm m} = 0.2758 \pm 0.0054$, which is in a $3.1\sigma$ tension with PP.  The LRA-PD model fit to the combination of Planck and DESI also does not reduce the tensions as well as predicted by the LRA, but still shows a significant reduction of it with $H_0 = 71.09 \pm 0.37$ km/s/Mpc, while improving both the Planck and the DESI $\chi^2$. It also yields $\Omega_{\rm m} = 0.2882 \pm 0.0045$ that is in a $2.5\sigma$ tension with PP. In both LRA models, the value of $S_8$ is lower than in the corresponding $\Lambda$CDM model. The impact of adding the PP dataset to the fit of the LRA-PD model is relatively minor, as shown in Fig.~\ref{fig:contours_all}, slightly shifting $\Omega_{\rm m}$ to a higher value and correspondingly lowering $H_0$ and increasing $S_8$. We note that the smallness of the cosmological parameter uncertainties in the LRA model fits is due to the fact that it has no additional free parameters -- the relative change in $x_e(z)$ is fixed.

The right panel of Fig.~\ref{fig:contours_all} shows the $H_0$, $S_8$ and  and $\Omega_{\rm m}$ posteriors from the four-parameter model fits to Planck+DESI, Planck+PPMb, and Planck+DESI+PPMb. The addition of the PP$M_b$ dataset plays the role of ``the SH0ES prior'', directing the fit towards ionization histories that yield larger values of $H_0$. To assess the goodness of these fits, Table~\ref{t:parameters_all} compares the Planck and the DESI $\chi^2$ to those the $\Lambda$CDM model. As one can see from Fig.~\ref{fig:contours_all} and Table~\ref{t:parameters_all}, a simple model of $x_e(z)$ can in fact relieve the $H_0$ tension while improving the CMB and BAO fits. When fit to Planck+DESI (without the SH0ES prior), it reduces the $H_0$ tension to $2.7\sigma$ while improving $\chi^2_{\rm Planck}$ by 3.3 and $\chi^2_{\rm BAO}$ by 2.8, a reduction in the total $\chi^2$ by $6.1$ after adding four parameters. When fit to Planck+PP$M_b$ and Planck+DESI+PP$M_b$, the four-parameter model yields $H_0$ values very similar to those from the corresponding LRA models.

The $x_e(z)$ corresponding to the Planck+DESI and Planck+DESI+PPMb best-fit 4-parameter models are shown in Fig.~\ref{fig:recom_fits} with the means and the 68\% CL uncertainties of the parameters $\Delta z_{\rm shift}$, $A_{\rm b}$, $z_{\rm b}$ and $\sigma_{\rm b}$ given in Table~\ref{t:parameters_all}. In addition to the expected $\Delta z_{\rm shift}<0$ favoring earlier recombination, the $x_e(z)$ in these models feature a dip in the $800 \lesssim z \lesssim 1100$ range, with a minimum around  $z\approx 930$. One can see that the Planck+DESI+PPMb best-fit four-parameter model $x_e(z)$ is broadly consistent with that of LRA-PD after smoothing over the wiggles.  

Based on the above results, one can conclude that the presence of wiggles in $x_e(z)$ is not essential for relieving the Hubble tension. To make this statement a bit more concrete, we compare the goodness of the fits more systematically in Table~\ref{t:delta_chi2_all} of Appendix~\ref{app:material}, where the differences in best fit $\chi^2$ are based on fits to identical data combinations (relative to the corresponding $\Lambda$CDM fit). Whereas the wiggly LRA model ionization histories perform somewhat better when fit to Planck + PP$M_b$, the same does not hold true when the DESI data are added. In particular, LRA-P Planck + PP$M_b$, with $\Delta \chi^2_{\rm total} = -34.77$, performs better than 4-par Planck + PP$M_b$, with $\Delta \chi^2_{\rm total} = -31.65$, while LRA-PD Planck + DESI + PP$M_b$, with $\Delta \chi^2_{\rm total} = -26.20$ perform slightly worse than 4-par Planck + DESI + PP$M_b$, with $\Delta \chi^2_{\rm total} = -27.18$. We note this comparison is not complete as it is based on a single LRA-predicted model, chosen based on a particular, not unique, combination of predicted $\chi^2_\mathrm{BF}$,  $H_0^\mathrm{BF}$, and $S_8^\mathrm{BF}$. It does, however, illustrate the point that a good fit can be achieved without the wiggles.

\FloatBarrier
\subsection{Implications for the high-$\ell$ CMB spectra}
\label{ss:highell}

\begin{table*}[!htbp]
	\begin{ruledtabular}
		\begin{tabular}{|c|ccc|cccc|}
			\multirow{2}{*}{\bf Parameter} &
			\boldmath$\Lambda\mathrm{CDM}$   & \bf{LRA-P}  & {\bf 4-par} Planck  & \boldmath$\Lambda\mathrm{CDM}$  & {\bf LRA-PD}  &{\bf 4-par} Planck  & {\bf 4-par} Planck \\
			& Planck          & Planck              & + PP$M_b$  & Planck + DESI       & Planck + DESI   &+ DESI     & + DESI + PP$M_b$ \\
			\colrule
			\rule{0pt}{3ex}
			$H_0$
			& \multirow{2}{*}{$67.24 \pm 0.47$} & \multirow{2}{*}{$72.57 \pm 0.47$} & \multirow{2}{*}{$72.21 \pm 0.89$} 
			& \multirow{2}{*}{$67.88 \pm 0.37$} & \multirow{2}{*}{$71.09 \pm 0.37$}  & \multirow{2}{*}{$69.23 \pm 0.98$} & \multirow{2}{*}{$71.34 \pm 0.68$} \\
			{[km/s/Mpc]} & & & & & & & \\[0.7ex]
			$H_0$ tension
			& $5.08\sigma$        & $0.41\sigma$        & $0.61\sigma$        & $4.67\sigma$      & $1.77\sigma$   & $2.67\sigma$     & $1.37\sigma$ \\
			$S_8$
			& $0.828 \pm 0.011$   & $0.795 \pm 0.010$ & $0.799 \pm 0.010$  & $0.815 \pm 0.009$  & $0.805 \pm 0.009$  & $0.812 \pm 0.010$ & $0.807 \pm 0.009$\\
			$S_8$ tension
			& $1.80\sigma$                & $0.24\sigma$    & $0.44\sigma$    & $1.24\sigma$          & $0.75\sigma$ & $1.07\sigma$ & $0.84\sigma$\\
			$\Omega_{\rm m}$
			& $0.315 \pm 0.006$  & $0.276 \pm 0.005$ & $0.279 \pm 0.006$ & $0.306 \pm 0.005$  & $0.288 \pm 0.004$ & $0.298 \pm 0.007$ & $0.286 \pm 0.005$\\
			$\Omega_m$ tension
			& $0.97\sigma$            & $3.10\sigma$  & $2.85\sigma$      & $1.47\sigma$               & $2.47\sigma$  & $1.85\sigma$ & $2.56\sigma$\\
			$\Omega_{\rm m} h^2$
			& $0.143 \pm 0.001$  & $0.145 \pm 0.001$ & $0.146 \pm 0.001$ & $0.141 \pm 0.001$  & $0.146 \pm 0.001$ & $0.143 \pm 0.002$ & $0.146 \pm 0.001$\\
			$100~\Omega_{\rm b} h^2$
			& $2.219 \pm 0.013$      & $2.283 \pm 0.013$  & $2.274 \pm 0.017$ & $2.229 \pm 0.013$      & $2.248 \pm 0.012$ & $2.246 \pm 0.018$ & $2.266 \pm 0.016$\\
			\colrule
			\rule{0pt}{3ex}
			$A_{\rm b}$
			& $-$                      & $-$    & $0.32 \pm 0.10$             & $-$      & $-$  &  $< 0.219$ & $0.30^{+0.10}_{-0.12}$\\[0.5ex]
			$z_{\rm b}$
			& $-$                         & $-$      & $936 \pm 22$           & $-$      & $-$  & $911^{+70}_{-21}$ & $928 \pm 22$\\[0.5ex]
			$\sigma_{\rm b}$
			& $-$                    & $-$  & $153^{+30}_{-20}$               & $-$      & $-$  & $< 194$  & $163^{+30}_{-20}$\\[0.5ex]
			$\Delta z_{\rm shift}$
			& $-$                & $-$  & $-43.1^{+7.6}_{-9.8}$               & $-$      & $-$  & $-19.5^{+13}_{-9.9}$ & $-39.1^{+7.7}_{-8.8}$\\[0.5ex]
			\colrule
			\rule{0pt}{3ex}
			$\chi^2_{\rm Planck}$
			& 10972.82                        & 10969.90  & 10972.82          & 10974.08         & 10973.50 & 10970.79 & 10972.87\\[0.5ex]
			$\chi^2_{\rm BAO}$
			& $-$                                  & $-$  & $-$               & 15.67            & 14.03    & 12.91    & 14.96
		\end{tabular}
	\end{ruledtabular}
	\caption{The mean values and the 68\% uncertainties for parameters of special interest from the MCMC runs represented in Fig.~\ref{fig:contours_all}, along with the best-fit Planck and DESI chi-squared values.}
	\label{t:parameters_all}
\end{table*}

As discussed in Sec.~\ref{s:methodology}, modified recombination relieves the Hubble tension by increasing $z_\star$, shifting the peak of the visibility function to an earlier epoch. However, changes to the shape of the visibility function that accompany such a shift are also very important, as they affect the amplitude of the E-mode polarization~\cite{Zaldarriaga:1995gi} and the suppression of CMB anisotropies in the Silk damping tail~\cite{Galli:2021mxk} at $\ell > 1500$. The high-$\ell$ anisotropy spectra, therefore, provide a critical test of these models in light of the high-resolution CMB experiments that are already placing nontrivial constraints on modified recombination~\cite{Thiele:2021okz,Galli:2021mxk}, and with more stringent tests expected with the data from the Simons Observatory~\cite{SimonsObservatory:2018koc} and CMB-S4~\cite{Abazajian:2019eic}.

\begin{figure}[!htbp]
	\centering
	\includegraphics[width=0.48\textwidth]{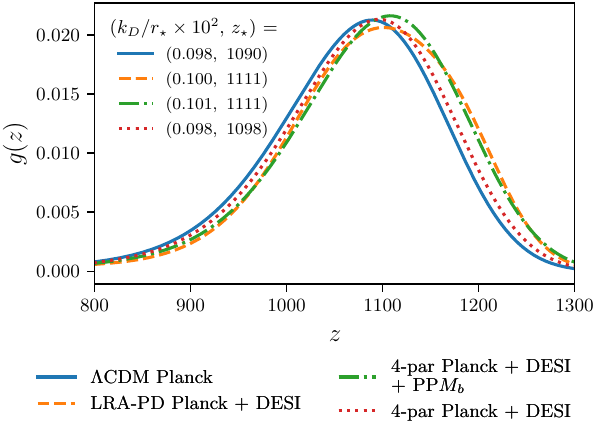}
	\caption{The visibility functions $g(z)$ for the Planck+DESI best-fit LRA-PD model, the Planck+DESI and Planck+DESI+PP$M_b$ best-fit four-parameter models, and the $\Lambda$CDM model. The legend lists the corresponding values of the ratio of the Silk damping scale to the sound horizon $k_D/r_\star \ [{\rm Mpc^{-2}}]$, and the redshift of the peak of the visibility function $z_\star$.}
	\label{fig:visibility}
\end{figure}

\begin{figure}[!htbp]
	\centering
	\includegraphics[width=0.48\textwidth]{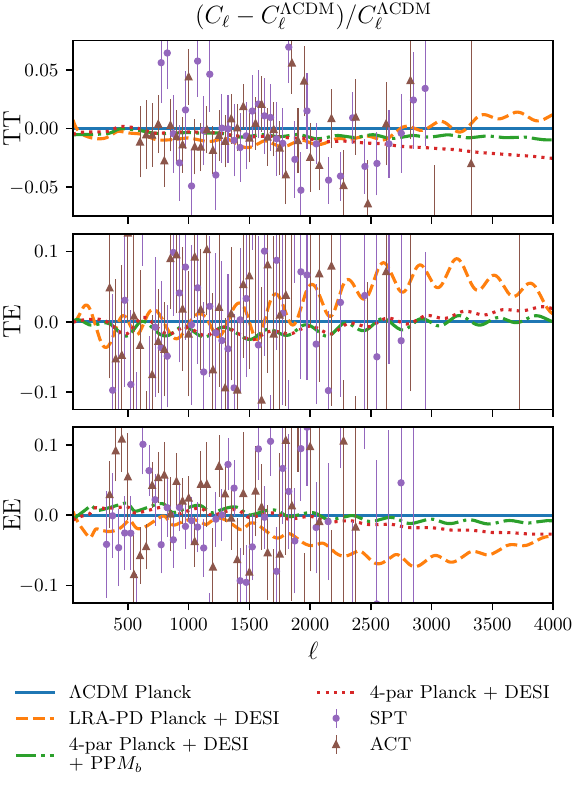}
	\caption{Relative differences in the CMB spectra with respect to the Planck best-fit $\Lambda$CDM model for the Planck+DESI best-fit LRA-PD model, and the Planck+DESI+PP$M_b$ and Planck+DESI best-fit four-parameter models. Also shown are the 2022 SPT-3G and ACT DR4 minimum-variance band power errors. In the case of TE, to avoid divisions by zero, we plot $(C^{\rm TE}_{\ell} - C_{\ell}^{{\rm TE}, \Lambda {\rm CDM}})/C_{\ell}^{\rm ref}$, where $C_{\ell}^{\rm ref}$ is the absolute value of $C_{\ell}^{{\rm TE}, \Lambda {\rm CDM}}$ convolved with a Gaussian of width $\sigma_{\ell} = 100$ centered at $\ell$.}
	\label{fig:cmb_residuals}
\end{figure}

\begin{table}[!htbp]
	\begin{ruledtabular}
		\begin{tabular}{|c|ccc|}
			\multirow{2}{*}{\bf Name} &
			{\bf 4-par} PL     & {\bf 4-par} PL      &{\bf 4-par} PL  \\
			& +DESI                    & +DESI +ACT            & +DESI +SPT     \\
			\colrule
			\rule{0pt}{3ex}
			$H_0$
			& $69.23 \pm 0.98 $ & $69.90 \pm 0.91$ & $69.11 \pm 0.91$ \\
			$S_8$
			& $0.812 \pm 0.010$   & $0.812 \pm 0.009$ & $0.811 \pm 0.010$  \\
			$\Omega_{\rm m}$
			& $0.298 \pm 0.007$  & $0.295 \pm 0.006$ & $0.299 \pm 0.007$ \\
			\colrule
			\rule{0pt}{3ex}
			$\Delta\chi^2_{\rm PL}$
			& $-3.28$ & $-5.34$ & $-1.50$ \\[0.5ex]
			$\Delta\chi^2_{\rm BAO}$
			& $-2.76$                               & $-1.25$  & $-2.89$  \\[0.5ex]
			$\Delta\chi^2_{\rm PL + BAO}$ 
			& $-6.04$                               & $-6.59$  & $-4.39$  \\[0.5ex]
			$\Delta\chi^2_{\rm ACT}$
			& $-$ & $-0.26$ & $-$ \\[0.5ex]
			$\Delta\chi^2_{\rm SPT}$
			& $-$ & $-$ & +2.92 \\[0.5ex]
			\colrule
			\rule{0pt}{3ex}
			$\Delta\chi^2_{\rm total}$
			& $-6.04$ & $-6.85$ & $-1.47$     
		\end{tabular}
	\end{ruledtabular}
	\caption{The mean values and the 68\% uncertainties for parameters of special interest along with the change in best-fit $\chi^2$ values relative the corresponding $\Lambda$CDM best fits for the four-parameter model fit to the combinations of Planck (denoted as PL), DESI, ACT, and SPT.}
	\label{t:parameters_sptact}
\end{table} 

Figure~\ref{fig:visibility} shows the visibility functions $g(z)$ corresponding to the best-fit recombination models discussed earlier. We see that the shifts in the peak are accompanied by different degree of broadening of the visibility function.\footnote{The visibility function is normalized to unity, so a lower peak indicates a broader function.} Notably, the LRA-PD model has a broader $g(z)$, while the two four-parameter models have $g(z)$ comparable or slightly narrower than of the $\Lambda$CDM model. We find that the ratio of the Silk damping scale to the sound horizon, \ie,~the ratio $k_D/r_\star$, remains nearly unchanged in modified recombination scenarios. Interestingly, this result has also been theoretically predicted in studies exploring variations of the electron mass to address the Hubble tension~\cite{Sekiguchi:2020teg, Schoneberg:2024ynd}. 

The corresponding relative differences in the CMB spectra at high $\ell$ are plotted in Fig.~\ref{fig:cmb_residuals} along with the ACT DR-4~\cite{ACT:2020gnv} and SPT-3G~\cite{SPT-3G:2022hvq} data. Interestingly, the LRA-PD model, which has the wider visibility function and wiggly $x_e(z)$, shows the larger differences with respect to $\Lambda$CDM at higher $\ell$, while the four-parameter Planck + DESI + PP$M_b$ spectra are generally close to those in $\Lambda$CDM despite achieving a slightly higher shift in $H_0$. We note that our LRA-PD CMB residuals are similar to those shown in Fig.~14 of Lynch \etal.~\cite{Lynch:2024gmp}. When fitting the LRA+PD model to Planck+DESI+ACT and Planck+DESI+SPT, we see a substantial increase in the ACT and the SPT $\chi^2$ compared to the corresponding $\Lambda$CDM fits, with the corresponding numbers provided in Tables~\ref{t:parameters_LCDM} and \ref{t:parameters_LRA} of Appendix~\ref{app:material}. One should keep in mind that the LRA-PD model was not optimized to fit ACT or STP, so such an outcome is not surprising. Nevertheless, it demonstrates how a model that provides a very good fit to Planck+DESI, while reducing the Hubble tension, can be strongly constrained even by the current high-$\ell$ anisotropy data.

As mentioned above, the high-$\ell$ anisotropy spectra are very sensitive not only to the peak but also to the shape of the visibility function. The spectra are affected by the combination of factors that include the Silk damping scale, which is generally smaller for earlier recombination, the amplitude of the $E$-mode polarization, which is generally larger for broader $g(z)$~\cite{Zaldarriaga:1995gi}, the optical depth $\tau$, the overall amplitude $A_s$, and the spectral index $n_s$. Our limited analysis in Sec.~\ref{ss:highell} does not point to a single dominant factor explaining the large difference in the residuals between the LRP-PD and the four-parameter Planck + DESI + PP$M_b$ model residuals, although it is expected that the parameter adjustments required to compensate for the notably broader $g(z)$ (which affects polarization differently from temperature anisotropies) in order to preserve the good fit to Planck, would yield larger differences at high $\ell$.

We also fit the four-parameter model to Planck+DESI+ACT and Planck+DESI+SPT to see how adding the high-$\ell$ datasets affects the ability of simple recombination models to resolve the Hubble tension. We perform these fits without the PPMb dataset, {\it i.e.} without the SH0ES prior. We find that, compared to fitting the four-parameter model to Planck+DESI, adding the ACT data yields a higher mean value of $H_0$, and further reducing the $\chi^2$ for all CMB datasets relative to the corresponding $\Lambda$CDM models, as shown in Table~\ref{t:parameters_sptact} (see also Table~\ref{t:parameters_4model} of Appendix~\ref{app:material}). On the other hand, adding the SPT data has the opposite effect, reducing the mean value of $H_0$ and increasing the $\chi^2$ values. This, again, demonstrates how important the future high-$\ell$ CMB anisotropy spectra will be for discriminating between modified recombination models.

\section{Discussion}
\label{s:discussion}

The results presented in the previous section are consistent with previous studies of modified recombination~\cite{Lee:2022gzh,Lynch:2024gmp,Lynch:2024hzh,Sekiguchi:2020teg,Schoneberg:2024ynd} and the general expectations for all models that attempt to relieve the Hubble tension by reducing the sound horizon. In particular, we have shown that it is possible to find ionization histories that increase the CMB-inferred value of $H_0$ to greater than $72$ km/s/Mpc, eliminating the tension between Planck and SH0ES. The existence of this possibility is not trivial as the successful models have to preserve the good fit to the fine features of the CMB TT, TE and EE spectra~\cite{Knox:2019rjx}. We found that the more flexible models of the ionization history based on cubic splines that allow for wiggly $x_e(z)$ were somewhat more successful in achieving high values of $H_0$ than the simple four-parameter model when fit to the Planck data alone. 

\begin{figure}[htbp]
	\centering
	\includegraphics[width=0.48\textwidth]{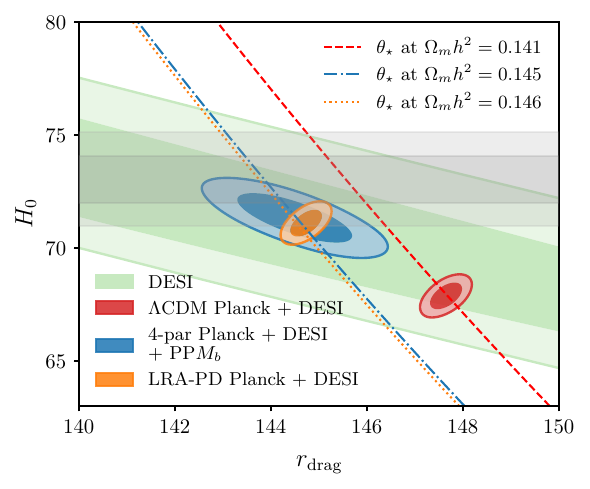}
	\caption{The 68\% and 95\% CL bands of $r_{\rm drag}$, $H_0$ derived from DESI BAO data marginalized over $\omega_m \equiv \Omega_m h^2$ (pale green), from the LRA-PD model (orange) fit to Planck+DESI, the four-parameter model (blue) fit to Planck+DESI+PPMb and the $\Lambda$CDM model fit to Planck+DESI (red). Also shown are the $r_{\rm drag}$-$H_0$ degeneracy lines defined by the CMB acoustic scale $\theta_\star$ at the $\Omega_mh^2$ values corresponding to the best-fit values for each model. The gray band shows the 68\% and 95\% CL determination of the Hubble constant by SH0ES.}
	\label{fig:desi_bao}
\end{figure}

Fully solving the Hubble tension while preserving a good fit to the combination of CMB and BAO is more challenging for the general reasons presented in \cite{Jedamzik:2020zmd} and which we briefly outline here. For a given value of $\Omega_m h^2$, the observed CMB acoustic scale $\theta_\star$ defines a line in the $r_{\rm drag}$-$H_0$ plane. Figure~\ref{fig:desi_bao} shows three such lines at three different values of $\Omega_m h^2$ corresponding to the best-fit $\Lambda$CDM, the best-fit LRA Planck+DESI and the Planck+DESI+PP$M_b$ best fit four-parameter model, as well as the 68\% and 95\% CL bands from DESI BAO after marginalizing over $\Omega_m h^2$, and the $H_0$ measurement by SH0ES. One can see that raising $H_0$ to values above $71$ km/s/Mpc by reducing $r_{\rm drag}$ is challenging, as it would require a much larger value of $\Omega_m h^2$ in order for the $\theta_\star$ line to pass through the region where the DESI and SH0ES bands overlap. Such large values of $\Omega_m h^2$ develop tension with galaxy weak lensing data from  DES and KiDs, and would worsen the $S_8$ tension~\cite{Jedamzik:2020zmd}.

With $H_0 > 71$ km/s/Mpc ruled out by the necessity to simultaneously accommodate the acoustic scales measured by both CMB and BAO, the need for the wiggly shape of $x_e(z)$ goes away. Our results suggest that even if the presence of wiggles in $x_e(z)$ could have been helpful for achieving a Planck-best-fit value of $H_0 > 72$ km/s/Mpc, it is not essential for getting $H_0 \sim 71$ km/s/Mpc. When combining Planck and DESI data, the 4-parameter model achieves a $\chi^2$ comparable to that of LRA-PD, while reaching an $H_0$ as high as $71$ km/s/Mpc. This point was also highlighted by Lynch \etal~\cite{Lynch:2024hzh} who demonstrated that a wiggly $x_e$ was not a necessity feature of the modified recombination models that accommodate a higher $H_0$. We also found that the Planck+DESI+PPMb best fit four-parameter models had smaller high-$\ell$ residuals relative to the Planck best-fit $\Lambda$CDM than the LRA-PD model.

For reference, Fig.~\ref{fig:recom_fits} also shows the ionization history obtained from magnetohydrodynamic (MHD) simulations of the primordial plasma through the epoch of recombination in the presence of PMF. The baryon clumping induced by PMF speeds up recombination and helps relieve the Hubble tension~\cite{Jedamzik:2020krr}. The $x_e(z)$ shown in Fig.~\ref{fig:recom_fits} is for a PMF with a blue Batchelor spectrum (predicted for PMF generated in phase transitions) and a root-mean-square post-recombination comoving strength of $B_{\rm rms} = 9.25\,$pG, and is based on recent MHD simulations that take into account all relevant effects, such as the Lyman-$\alpha$ photon transport across regions of higher and lower density~\cite{Jedamzik:2023rfd}. It is intriguing that this $x_e(z)$, which provides a good fit to the combination of Planck and DESI, has a shape that is not too far from the ionization histories of the best fit four-parameter models. This is particularly nontrivial given the very low success rate when searching for viable modified ionization histories using the LRA method, as mentioned at the beginning of Sec~\ref{s:results}.

The $S_8$ tensions is trivially reduced in models of modified recombination that yield larger values of $H_0$. This is because $\Omega_m h^2$ does not change by much in these models, and neither do the other cosmological parameters that determine $\sigma_8$. Hence, a larger $h$ automatically implies a smaller $\Omega_m$ and hence a smaller $S_8$ ($S_8 \equiv \sigma_8 \sqrt{\Omega_m /0.3}$). 

The fact that modified recombination solutions to the Hubble tension generically reduce $\Omega_m$ exacerbates the disagreement with the larger $\Omega_m$ measured from the recent SN datasets, as also pointed out in~\cite{Lee:2022gzh,Baryakhtar:2024rky,Poulin:2024ken}. However, for the models we studied, this tension is in the 2$-$3$\sigma$ range (see Table~\ref{t:parameters_all}), which is relatively mild. Furthermore, while a reliable determination of $\Omega_m$ from uncalibrated BAO and SN is of direct relevance to the Hubble tension discussion~\cite{Blanchard:2022xkk,Poulin:2024ken,Pedrotti:2024kpn},
there are good reason for viewing the root cause of the $\Omega_m$ problem as being separate from the Hubble tension. It arises from the fact that the shapes of the conformal distance-redshift curves deduced from the BAO and SN datasets covering the same redshift range are not well accommodated by the same flat $\Lambda$CDM cosmology -- DESI BAO measure $\Omega_m = 0.295 \pm 0.015$~\cite{DESI:2024mwx}, while PP SN yield $\Omega_m = 0.334 \pm 0.018$~\cite{Brout:2022vxf}. Departures from the flat $\Lambda$CDM, such as the dynamical dark energy models which fit the BAO+SN combination better, do not help raise the value of $H_0$~\cite{DESI:2024mwx}. We also note that Planck and DESI are in excellent consistency with each other in models with lower sound horizon that yield $H_0 \sim 70$ km/s/Mpc~\cite{Pogosian:2024ykm}.

\section{Summary}
\label{s:summary}

Our study demonstrated that it is possible to reduce the Hubble tension below 2$\sigma$ by modifying the ionization history $x_e(z)$, while preserving and even improving the fit to the current CMB and BAO data and reducing the $S_8$ tension. We considered two models of $x_e(z)$: a fully agnostic parametrization using a cubic-spline through 7 nodes in redshift, and a phenomenological four-parameter model motivated by the ionization histories derived from simulations of recombination in the presence of PMF. While the cubic-spline parametrization yields wiggly best-fit shapes of $x_e(z)$, our analysis confirmed that the presence of wiggles is not essential for achieving good fits to the combination of CMB and BAO data with $H_0 \sim 71$ km/s/Mpc. Our study also confirmed that the high-resolution CMB temperature and polarization anisotropies measured by ACT, SPT, Simons Observatory~\cite{SimonsObservatory:2018koc} and CMB-S4~\cite{Abazajian:2019eic} have a crucial role in discriminating between models of modified recombination. Overall, modifying recombination remains a viable option for resolving the Hubble tension.

\acknowledgments

We thank Jens Chluba, Gabriel Lynch, Ali Nezhadsafavi, and Xavier Wang for useful discussions. This research was enabled in part by support provided by the BC DRI Group and the Digital Research Alliance of Canada ({\tt alliancecan.ca}). S.H.M and L.P. are supported in part by the National Sciences and Engineering Research Council (NSERC) of Canada.

\section{Data availability}
\label{s:data_avail}

The data supporting this study's findings are available within the article.

\appendix
\section{Details of the LRA analysis}
\label{app:numerics}

In this section, we provide additional details related to the LRA analysis, where we closely followed the methods used in \cite{Lee:2022gzh}. 

\subsection{Calculating the shifts in the best-fit parameters}
\label{ss:shifts}

Assuming the modified recombination model is sufficiently close to the fiducial $\Lambda$CDM cosmology, one can compute the change in the observable $\mathbi{X}$ due to a perturbation of the ionization history $\Delta \mathrm{ln} \, [x_e(z)]$ as
\be
\label{eq:DX}
\Delta \mathbi{X} = \int dz\; \frac{\delta \mathbi{X}}{\delta \mathrm{ln} \, [x_e(z)]} \Delta \mathrm{ln} \, [x_e(z)].
\ee
The best-fit parameters and chi-squared of the modified model will be different from those of the fiducial model. The shifts in the best-fit parameters and chi-squared due to a perturbation $\Delta \mathrm{ln} \, [x_e(z)]$ can be written as
\begin{subequations}
	\label{eq:shift_bf}
	\begin{align}
	&\Delta \Omega_{\rm BF}^i = \int dz\; \frac{\delta \Omega_{\rm BF}^i}{\delta \mathrm{ln} \, [x_e(z)]} \Delta \mathrm{ln} \, [x_e(z)]
	,\label{eq:DObf_X} \\
	\begin{split}
		&\Delta \chi_{\rm BF}^2 = \int dz\; \frac{\delta \chi_{\rm BF}^2}{\delta \mathrm{ln} \, [x_e(z)]}\Delta \mathrm{ln} \, [x_e(z)] \\
		&+ \frac12 \iint dz \, dz' \frac{\delta^2 \chi_{\rm BF}^2}{\delta \mathrm{ln} \, [x_e(z)] \delta \mathrm{ln} \, [x_e(z')]} \Delta \mathrm{ln} \, 	[x_e(z)] \Delta \mathrm{ln} \, [x_e(z')],
		\label{eq:Dchi2bf_X}
	\end{split}
	\end{align}
\end{subequations}
where
\begin{subequations}
	\label{eq:deriv_bf}
	\begin{align}
		&\frac{\delta \Omega_{\rm BF}^i}{\delta \mathrm{ln} \, [x_e(z)]} = -(F^{-1})_{ij} \frac{\partial \mathbi{X}}{\partial \Omega^j} \cdot \mathbi{M}\cdot\frac{\delta \mathbi{X}}{\delta \mathrm{ln} \, [x_e(z)]},\label{eq:dObf}\\
		&\frac{\delta \chi_{\rm BF}^2}{\delta \mathrm{ln} \, [x_e(z)]} = 2[\mathbi{X}(\vec{\Omega}_{\rm fid}) - \bm{X}^{\rm obs}] \cdot\widetilde{\boldsymbol{M}} \cdot \frac{\delta \mathbi{X}}{\delta \mathrm{ln} \, [x_e(z)]},\label{eq:dchi2bf_lin}\\
		&\frac{\delta^2 \chi_{\rm BF}^2}{\delta \mathrm{ln} \, [x_e(z)] \delta \mathrm{ln} \, [x_e(z')]} = 2 \frac{\delta \mathbi{X}}{\delta \mathrm{ln} \, [x_e(z)]}\cdot\widetilde{\boldsymbol{M}}\cdot\frac{\delta \mathbi{X}}{\delta \mathrm{ln} \, [x_e(z')]},\label{eq:dchi2bf_quad}
	\end{align}
\end{subequations}
where $F_{ij}, \mathbi{M}, \widetilde{\boldsymbol{M}}$, and $\delta \mathbi{X}/\delta \mathrm{ln} \, [x_e(z)]$ are all to be evaluated at the fiducial cosmology and in the standard model. Here, $\boldsymbol{\widetilde{M}}$ is defined as
\be
\label{eq:tilde_M}
\widetilde{M}_{\alpha\beta} \equiv M_{\alpha\beta} - M_{\alpha \gamma} \frac{\partial X^\gamma}{\partial \Omega^i}(F^{-1})_{ij}\frac{\partial X^\sigma}{\partial \Omega^j}M_{\sigma\beta},
\ee
and it can be interpreted as the inverse of the covariance matrix after marginalization over shifts in standard cosmological parameters~\cite{Lee:2022gzh}.

\subsection{Evaluating functional derivatives}

\begin{figure}[htbp]
	\centering
	\includegraphics[width=0.48\textwidth]{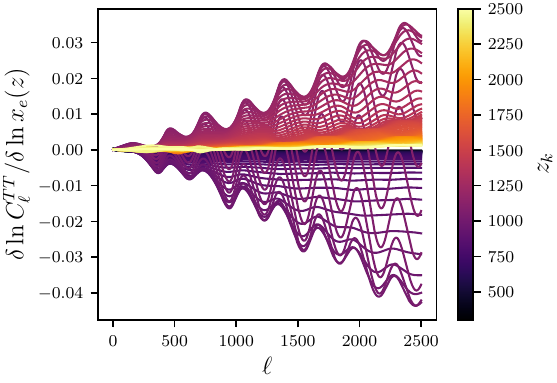}  
	\caption{Derivatives of the CMB temperature power spectrum with respect to the ionized fraction evaluated at a discrete set of redshifts $z_k$ spanning the range $[z_{\rm min}, z_{\rm max}] = [300, 2500]$, using the basis functions in (\ref{eq:basis_funcs}).}
	\label{fig:dcldlogx}
\end{figure}

\begin{figure*}[htbp]
	\centering
	\includegraphics[width=0.9\textwidth]{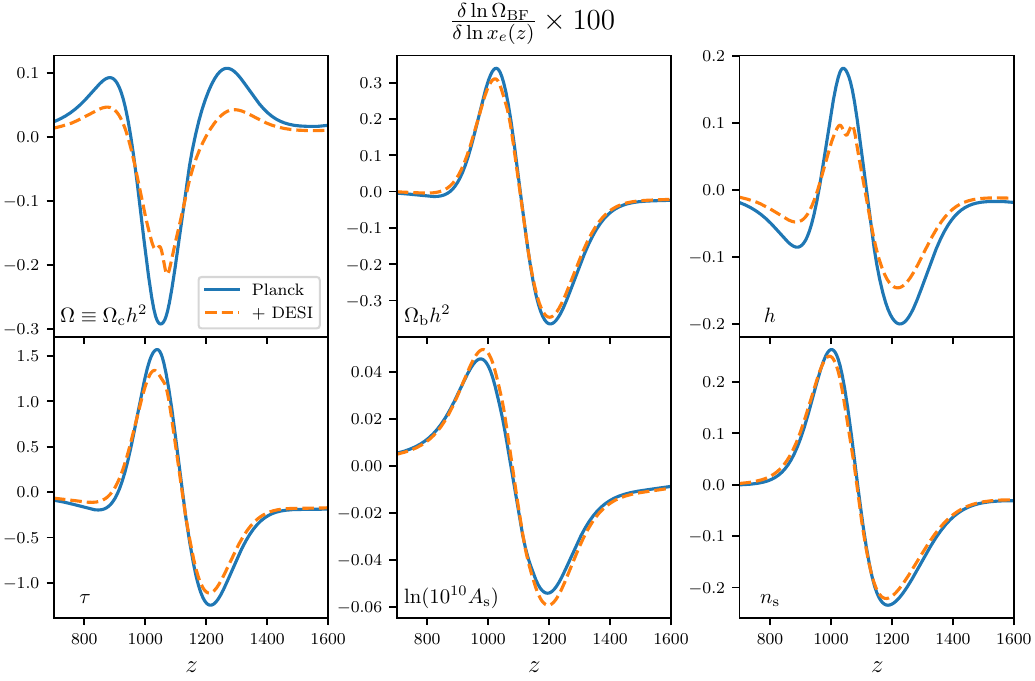}
	\caption{Functional derivatives of the best-fit cosmological parameters for Planck and Planck+DESI data. The plots cover the redshift range of $z = [700, 1600]$. The CMB data show little sensitivity to perturbations in $x_e(z)$ outside this range.}
	\label{fig:domegadf_all}
\end{figure*}

\begin{figure}[htbp]
		\includegraphics[width=0.45\textwidth]{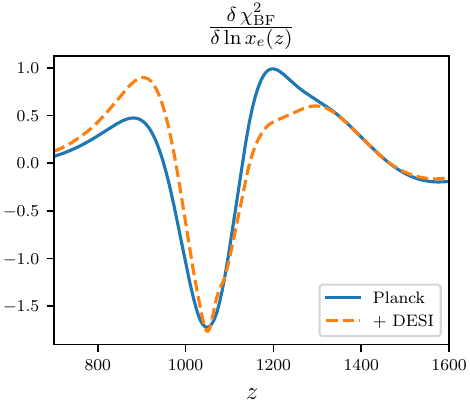}
\caption{The functional derivatives of the best-fit chi-squared for Planck and Planck+ DESI data.}
\label{fig:dchi2df_all}
\end{figure}

\begin{figure*}[htbp]
	\centering
	\begin{subfigure} 
		\centering
		\includegraphics[width=0.45\textwidth]{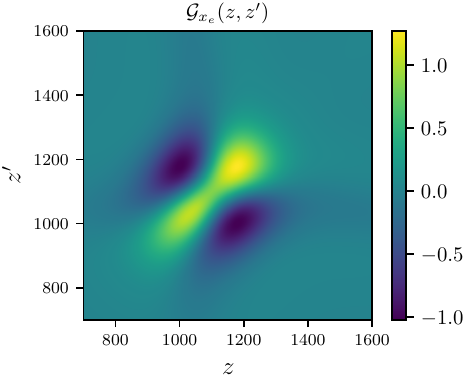} 
	\label{fig:d2chi2df2}
	\end{subfigure}
	\hspace{0.2cm}
	\begin{subfigure} 
		\centering
		\includegraphics[width=0.45\textwidth]{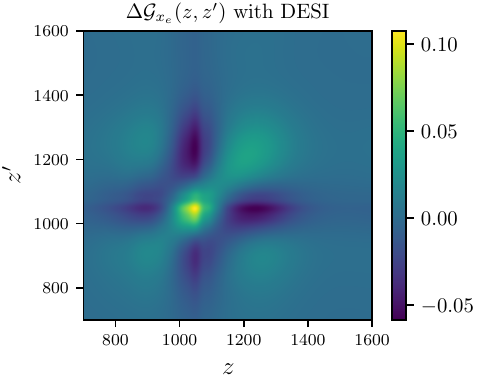}
		\label{fig:d2chi2df2_all}
\end{subfigure}
	\caption{The quadratic response of the best-fit chi-squared $\mathcal{G}_{x_e}(z, z') \equiv \delta^2 \chi^2_{\mathrm{BF}} / \delta \mathrm{ln} \, [x_e(z)] \, \delta \mathrm{ln} \, [x_e(z')]$ with respect to logarithmic variations in $x_e(z)$ at each redshift for the Planck data (left) and DESI BAO data (right).}
	\label{fig:dchi2_multi}
\end{figure*}

We build an orthonormal set of response functions to form a vector space that generates perturbations in $x_e(z)$ over given narrow redshift ranges. We define Gaussian shapes used for the principal component analysis in \cite{Hart:2019gvj}, since they are smooth and easy to implement into the Boltzmann equation solver.\footnote{Strictly speaking, Gaussian functions are not orthogonal to each other, but for all practical purposes, they are when hardly overlapping.} 
The generated basis functions are given by
\be
\label{eq:basis_funcs}
	\begin{split}
		&\Delta \mathrm{ln} \, [x_e(z, z_k)] \equiv \f{\Delta x_e}{x_0} (z, z_k) \\
		& = \f{1}{\sqrt{2\pi \sigma^2}} \mathrm{exp} \l[ - \f{(z - z_{k})^2}{2 \sigma^2} \r], \quad \sigma \equiv \f{z_{max} - z_{min}}{6 N \sqrt{2 \mathrm{ln 2}}}.
	\end{split}
\ee
Here, $z_{\rm min}$ and $z_{\rm max}$ define the boundaries of the redshift range, while $N$ represents the number of basis functions, each associated with a redshift $z_k$ within that range. We generate $N = 121$ basis functions in the redshift range of $[z_{\rm min}, z_{\rm max}] = [300, 2500]$. This selection of $N$ guarantees the convergence of sampled components, being large enough to remove all nonorthogonalities, yet not too large to cause highly nonlinear responses where obtained functions are no longer smooth. The selected redshift range covers the epoch of hydrogen and helium recombination. Outside this range, the CMB anisotropies vary negligibly unless we move into the reionization era at $z \le 10$.

We implement the basis functions via {\tt RECFAST} into the Boltzmann solver code {\tt CAMB}, and analyze the response of the CMB anisotropy power spectra, and BAO measurements. The functional derivatives of the observables at each redshift $z_k$ are calculated via numerical differentiation by using three-point midpoint formula:
\begin{equation}
\label{eq:func_deriv}
	\begin{split}
		&\f{\delta{\mathbi{X}}}{\delta{\mathrm{ln} x_e(z)}} \equiv \f{\delta{\mathbi{X}}}{\delta{x_e(z)/x_0}} \\
		&= \f{\mathbi{X}[x_0 + \Delta x_e(z, z_k)] - \mathbi{X}[x_0 - \Delta x_e(z, z_k)]}{2 \Delta x_e(z, z_k)/x_0} \bigg|_{z = z_k}.
	\end{split}
\end{equation}
The accuracy of these derivatives is controlled by the parameter $N$, which sets the widths of and the intervals between the Gaussians. To verify the convergence of functional derivatives, we varied $N$, and examined the convergence of numerically calculated derivatives at 23 selected redshifts between $300$ and $2500$. We found that choosing $N < 200$ ensured convergence across all the redshift points. However, selecting $N > 140$ introduced unwanted noise due to overlaps in the sampled components. A choice of $N=121$ provided an optimal balance: it was large enough to ensure convergence, yet not too large as to introduce nonlinear responses.

The changes in the ionization history affect CMB anisotropies via the Thomson visibility function. Figure~\ref{fig:dcldlogx} shows the derivative of the logarithm of the temperature power spectrum with respect to logarithm of $x_e$ at redshifts $z_k$. The strongest response occurs around the peak of the visibility function ($z_* \simeq 1100$) with about 4\% maximum amplitude. The color bar shows that the change in the slopes of spectra is positive for $z > z_*$, and a negative response for $z < z_*$. The precise calculation of the functional derivatives requires a boost in the {\tt CAMB} accuracy parameters. In particular, we reduced the time step by a factor of 20, as one needs a high resolution time sampling around recombination to avoid discontinuities in the Boltzmann equation solver, achieving numerically stable responses.

With the derivatives of the observables at hand, we use (\ref{eq:deriv_bf}) to calculate the functional derivatives of the cosmological parameters and chi-squared, which are then used to find the shifts in the best-fit values. Figures \ref{fig:domegadf_all}, \ref{fig:dchi2df_all} and \ref{fig:dchi2_multi} illustrate these derivatives, showing the changes in the best-fit cosmological parameters, best-fit chi-squared, and the quadratic response of the best-fit chi-squared $\delta^2 \chi^2_{\mathrm{BF}} / \delta \mathrm{ln} \, [x_e(z)] \delta \mathrm{ln} \, [x_e(z')]$ with respect to variations in $\mathrm{ln} [x_e(z)]$.

\subsection{Generating trial recombination histories $\Delta \mathrm{ln} \, [x_e(z)]$}

We parametrize $\Delta \mathrm{ln} \, [x_e(z)] = \Delta x_e(z)/x_0$ using a cubic spline passing through 7 control points placed evenly between $z_a = 700$ and $z_b = 1600$. We set $\Delta x_e = 0$ outside this range, as the CMB data show negligible sensitivity to the changes in ionization history beyond these redshifts, as illustrated in Fig. \ref{fig:domegadf_all}. The departures from standard recombination at each control point are chosen from 6 amplitudes uniformly spaced between 0 and $dx_{\rm max}$. This configuration provides us with the possibility of generating a large number (i.e., $6^7\approx 280000$) of recombination histories to investigate deviations from the standard recombination. We also tried four different values of $dx_{\rm max}$, $(-0.2,-0.25,-0.3,-0.35)$. We developed a PYTHON pipeline to identify candidate splines that meet our criteria of $H_0^{\rm BF} \ge 70$, $\Delta \chi^2_{\rm BF} \le 0$. Using a modified version of CAMB that takes the generated cubic splines and the corresponding best-fit cosmological parameters from the PYTHON code as input, we calculate the derived value of $S_8$ parameter. This allows us to apply our third criterion, $S_8^{\rm BF} (\Delta \mathrm{ln} \, [x_e(z)]) < S_{8}^{\rm fid}$, to filter out the remaining ineligible splines. 

\onecolumngrid
\section{Parameter tables}
\label{app:material}

\begin{table*}[htbp]
	\begin{ruledtabular}
		\begin{tabular}{|c|cccc|ccc|}
			\multirow{2}{*}{\boldmath $\Delta \chi^2_{\rm BF}$} &
			{\bf LRA-P} & {\bf LRA-PD} & {\bf LRA-P} & {\bf LRA-PD} Planck & {\bf 4-par} Planck & {\bf 4-par} Planck & {\bf 4-par} Planck \\
			& Planck & Planck + DESI & Planck + PP$M_b$ & + DESI + PP$M_b$  &  +DESI & + PP$M_b$ & + DESI + PP$M_b$ \\
			\colrule
			\rule{0pt}{3ex}
			low-$\ell$ TT       & +0.36              & +3.68              & +1.67     & +3.44       & +0.10        & +1.61              & +0.37 \\
			low-$\ell$ EE       & +0.05              & $-0.08$            & $-0.68$   & $-1.15$     & $-0.10$      & $-0.82$            & $-1.18$ \\
			high-$\ell$         & $-4.74$            & $-4.07$            & $-9.78$   & $-5.69$     & $-2.96$      & $-6.08$            & $-3.21$ \\
			lensing             & +1.41              & $-0.10$            & $-0.69$   & $-0.14$     & $-0.33$      & $-1.02$            & $-0.51$ \\
			\colrule
			\rule{0pt}{3ex}
			$\Delta \chi^2_{\rm CMB}$
			& $-2.92$            & ${-0.58}$           & ${-9.49}$ & ${-3.55}$   & $-3.29$     & ${-6.30}$          & ${-4.53}$ \\[0.5ex]
			$\Delta \chi^2_{\rm BAO}$
			& $-$                & $-1.64$            & $-$       & +0.53        & $-2.76$     & $-$                & +1.42 \\[0.5ex]
			$\Delta \chi^2_{\rm SN}$
			& $-$                & $-$                & +4.33     & +2.85        & $-$         & +5.06              & +3.90 \\[0.5ex]
			$\Delta \chi^2_{\rm SH0ES}$
			& $-$                & $-$                & $-29.61$  & $-26.03$     & $-$         & $-30.41$           & $-27.97$ \\[0.5ex]
			\colrule
			\rule{0pt}{3ex}
			$\Delta \chi^2_{\rm total}$
			& $-2.92$            & $-2.22$            & $-34.77$  & $-26.20$     & $-6.04$     & $-31.65$           & $-27.18$
		\end{tabular}
	\end{ruledtabular}
	\caption{Changes in the best-fit chi-squared values $\Delta \chi^2_{\rm BF}$ for the LRA-P, LRA-PD and the four-parameter model fits relative to the $\Lambda$CDM fits to the same data combinations.}
	\label{t:delta_chi2_all}
\end{table*}

\begin{table*}[htbp]
	\begin{ruledtabular}
		\begin{tabular}{|c|cccccc|}
			\multirow{2}{*}{\bf Parameter}
			& \boldmath$\Lambda\mathrm{CDM}$  & \boldmath$\Lambda\mathrm{CDM}$  & \boldmath$\Lambda\mathrm{CDM}$ Planck
			& \boldmath$\Lambda\mathrm{CDM}$ Planck  & \boldmath$\Lambda\mathrm{CDM}$ Planck  & \boldmath$\Lambda\mathrm{CDM}$ Planck \\
			& Planck           & Planck + DESI        & + DESI + ACT
			& + DESI + SPT       & + DESI + PP  & + DESI + PP$M_b$  \\
			\colrule
			\rule{0pt}{3ex}
			$H_0$ [km/s/Mpc]
			& $67.24 \pm 0.47$ & $67.88 \pm 0.37$ & $67.97 \pm 0.35$ & $67.87 \pm 0.35$ & $67.77 \pm 0.36$ & $68.50 \pm 0.34$ \\[0.5ex]
			$H_0$ tension
			& $5.08\sigma$ & $4.67\sigma$ & $4.62\sigma$ & $4.71\sigma$ & $4.79\sigma$ & $4.15\sigma$ \\[0.5ex]
			$S_8$
			& $0.828 \pm 0.011$ & $0.815 \pm 0.009$ & $0.817 \pm 0.009$ & $0.814 \pm 0.009$ & $0.817 \pm 0.009$ & $0.804 \pm 0.009$ \\[0.5ex]
			$S_8$ tension
			& $1.80\sigma$ & $1.24\sigma$ & $1.34\sigma$ & $1.19\sigma$ & $1.34\sigma$ & $0.70\sigma$ \\[0.5ex]
			$\Omega_{\rm m}$
			& $0.315 \pm 0.006$ & $0.306 \pm 0.005$ & $0.305 \pm 0.005$ & $0.306 \pm 0.005$ & $0.308 \pm 0.005$ & $0.299 \pm 0.004$ \\[0.5ex]
			$\Omega_{\rm m}$ tension
			& $0.97\sigma$ & $1.47\sigma$ & $1.55\sigma$ & $1.47\sigma$ & $1.39\sigma$ & $1.90\sigma$ \\[0.5ex]
			$\Omega_{\rm m} h^2$
			& $0.143 \pm 0.001$ & $0.141 \pm 0.001$ & $0.141 \pm 0.001$ & $0.141 \pm 0.001$ & $0.141 \pm 0.001$ & $0.140 \pm 0.001$ \\[0.5ex]
			$\tau_{\rm reio}$
			& $0.052 \pm 0.007$ & $0.057 \pm 0.007$ & $0.056 \pm 0.007$ & $0.055 \pm 0.007$ & $0.056 \pm 0.007$ & $0.062^{+0.007}_{-0.008}$ \\[0.5ex]
			$n_{\rm s}$
			& $0.964 \pm 0.004$ & $0.967 \pm 0.004$ & $0.970 \pm 0.003$ & $0.967 \pm 0.003$ & $0.967 \pm 0.004$ & $0.970 \pm 0.003$ \\[0.5ex]
			$\mathrm{log}(10^{10} A_{\rm s})$
			& $3.037 \pm 0.014$ & $3.044 \pm 0.014$ & $3.051 \pm 0.013$ & $3.042 \pm 0.013$ & $3.043 \pm 0.014$ & $3.052^{+0.014}_{-0.015}$ \\[0.5ex]
			$M_b$
			& $-$ & $-$ & $-$ & $-$ & $-$ & $-19.406 \pm 0.010$ \\[0.5ex]
			$r_\star$ [Mpc]
			& $144.65 \pm 0.24$ & $144.95 \pm 0.20$ & $144.97 \pm 0.20$ & $144.94 \pm 0.19$ & $144.89 \pm 0.19$ & $145.16 \pm 0.19$ \\[0.5ex]
			$r_{\rm drag}$ [Mpc]
			& $147.38 \pm 0.24$ & $147.65 \pm 0.21$ & $147.67 \pm 0.21$ & $147.67 \pm 0.21$ & $147.60 \pm 0.21$ & $147.81 \pm 0.21$ \\[0.5ex]
			$r_{\rm drag} h$ [Mpc]
			& $99.10 \pm 0.81$ & $100.23 \pm 0.64$ & $100.36 \pm 0.62$ & $100.16 \pm 0.62$ & $100.01 \pm 0.60$ & $101.26 \pm 0.58$ \\[0.5ex]
			$z_\star$
			& $1090.14 \pm 0.23$ & $1089.89 \pm 0.20$ & $1089.88 \pm 0.18$ & $1089.90 \pm 0.18$ & $1089.94 \pm 0.19$ & $1089.61 \pm 0.18$ \\[0.5ex]
			$z_{\rm drag}$
			& $1059.51 \pm 0.28$ & $1059.63 \pm 0.28$ & $1059.63 \pm 0.25$ & $1059.63 \pm 0.25$ & $1059.61 \pm 0.27$ & $1059.88 \pm 0.27$ \\[0.5ex]
			$10^4\theta_\star$
			& $104.098 \pm 0.025$ & $104.115 \pm 0.024$ & $104.127 \pm 0.022$ & $104.106 \pm 0.022$ & $104.112 \pm 0.024$ & $104.132 \pm 0.023$ \\[0.5ex]
			$100Y_{\rm P}$
			& $24.578 \pm 0.006$ & $24.582 \pm 0.005$ & $24.582 \pm 0.005$ & $24.582 \pm 0.005$ & $24.581 \pm 0.005$ & $24.588 \pm 0.005$ \\[0.5ex]
			$\sigma_8$
			& $0.808 \pm 0.005$ & $0.806 \pm 0.005$ & $0.810 \pm 0.005$ & $0.805 \pm 0.005$ & $0.807 \pm 0.006$ & $0.806 \pm 0.006$ \\[0.5ex]
			\colrule
			\rule{0pt}{3ex}
			$\chi^2_{\rm low\text{-}\ell\ TT}$
			& 23.80 & 22.74 & 22.16 & 22.55 & 23.48 & 22.83 \\[0.5ex]
			$\chi^2_{\rm low\text{-}\ell\ EE}$
			& 395.85 & 396.52 & 397.45 & 395.90 & 396.84 & 396.82 \\[0.5ex]
			$\chi^2_{\rm high\text{-}\ell}$
			& 10544.46 & 10545.91 & 10549.43 & 10549.55 & 10544.97 & 10548.23 \\[0.5ex]
			$\chi^2_{\rm lensing}$
			& 8.71 & 8.91 & 8.37 & 9.24 & 8.76 & 9.53 \\[0.5ex]
			\colrule
			\rule{0pt}{3ex}
			$\chi^2_{\rm Planck}$
			& 10972.82 & 10974.08 & 10977.41 & 10977.24 & 10974.05 & 10977.41 \\[0.5ex]
			$\chi^2_{\rm BAO}$
			& $-$ & 15.67 & 14.34 & 16.59 & 16.22 & 13.54 \\[0.5ex]
			$\chi^2_{\rm SN}$
			& $-$ & $-$ & $-$ & $-$ & 1404.68 & 1406.35 \\[0.5ex]
			$\chi^2_{\rm SH0ES}$
			& $-$ & $-$ & $-$ & $-$ & $-$ & 33.99 \\[0.5ex]
			$\chi^2_{\rm ACT}$
			& $-$ & $-$ & 239.73 & $-$ & $-$ & $-$ \\[0.5ex]
			$\chi^2_{\rm SPT}$
			& $-$ & $-$ & $-$ & 1876.88 & $-$ & $-$ \\[0.5ex]
			\colrule
			\rule{0pt}{3ex}
			$\chi^2_{\rm total}$
			& 10972.82 & 10989.75 & 11231.48 & 12870.70 & 12394.95 & 12431.30
		\end{tabular}
	\end{ruledtabular}
	\caption{Mean values and the 68\% CL intervals of the parameters and the $\chi^2_{\rm BF}$ values for the $\Lambda$CDM model fit to combinations of Planck, DESI, ACT, SPT, PP, and PP$M_b$.}
	\label{t:parameters_LCDM}
\end{table*}

\begin{table*}[htbp]
	\begin{ruledtabular}
		\begin{tabular}{|c|cc|cccc|}
			\multirow{3}{*}{\bf Parameter} &
			\multicolumn{2}{c|}{\bf{LRA-P}} & \multicolumn{4}{c|}{\bf{LRA-PD}} \\
			\cline{2-3}  \cline{4-7}			
			\rule{0pt}{3ex}
			& \multirow{2}{*}{Planck}              & {Planck}    & Planck    & Planck + DESI   & Planck + DESI   & Planck + DESI \\
			&  & + PP & + DESI & + ACT & + SPT & + PP \\
			\colrule
			\rule{0pt}{3ex}
			$H_0$ [km/s/Mpc]
			& $72.57 \pm 0.47$ & $72.12 \pm 0.45$ & $71.09 \pm 0.37$ & $71.2 \pm 0.36$ & $71.14 \pm 0.35$ & $70.89 \pm 0.36$ \\[0.5ex]
			$H_0$ tension
			& $0.41\sigma$ & $0.81\sigma$ & $1.77\sigma$ & $1.67\sigma$ & $1.73\sigma$ & $1.95\sigma$ \\[0.5ex]
			$S_8$
			& $0.795 \pm 0.010$ & $0.8033 \pm 0.0099$ & $0.8055 \pm 0.0090$ & $0.8064 \pm 0.0089$ & $0.8035 \pm 0.0086$ & $0.8095 \pm 0.0088$ \\[0.5ex]
			$S_8$ tension
			& $0.24\sigma$ & $0.65$ & $0.75\sigma$ & $0.82\sigma$ & $0.68\sigma$ & $0.97\sigma$ \\[0.5ex]
			$\Omega_{\rm m}$
			& $0.2758 \pm 0.0054$ & $0.2810 \pm 0.0052$ & $0.2882 \pm 0.0045$ & $0.2870 \pm 0.0043$ & $0.2876 \pm 0.0043$ & $0.2908 \pm 0.0044$ \\[0.5ex]
			$\Omega_{\rm m}$ tension
			& $3.10\sigma$ & $2.83\sigma$ & $2.47\sigma$ & $2.54\sigma$ & $2.51\sigma$ & $2.33\sigma$\\[0.5ex]
			$\Omega_{\rm m} h^2$
			& $0.1452 \pm 0.0010$ & $0.1461 \pm 0.0010$ & $0.1457 \pm 0.0008$ & $0.1455 \pm 0.0008$ & $0.1455 \pm 0.0008$ & $0.1461 \pm 0.0008$ \\[0.5ex]
			$\tau_{\rm reio}$
			& $0.0558 \pm 0.0072$ & $0.0528 \pm 0.0071$ & $0.0515 \pm 0.0068$ & $0.0501 \pm 0.0067$ & $0.0499 \pm 0.0067$ & $0.0500 \pm 0.0067$ \\[0.5ex]
			$n_{\rm s}$
			& $0.9620 \pm 0.0038$ & $0.9597 \pm 0.0038$ & $0.9518 \pm 0.0036$ & $0.9553 \pm 0.0032$ & $0.0499 \pm 0.0067$ & $0.0500 \pm 0.0067$ \\[0.5ex]
			$\mathrm{log}(10^{10} A_{\rm s})$
			& $3.042 \pm 0.014$ & $3.038 \pm 0.014$ & $3.028 \pm 0.013$ & $3.034 \pm 0.013$ & $3.027 \pm 0.013$ & $3.026 \pm 0.013$ \\[0.5ex]
			$r_\star$ [Mpc]
			& $141.52 \pm 0.23$ & $141.32 \pm 0.22$ & $141.92 \pm 0.20$ & $141.97 \pm 0.19$ & $141.95 \pm 0.19$ & $141.83 \pm 0.19$ \\[0.5ex]
			$r_{\rm drag}$ [Mpc]
			& $144.20 \pm 0.24$ & $144.03 \pm 0.23$ & $144.74 \pm 0.21$ & $144.78 \pm 0.21$ & $144.76 \pm 0.20$ & $144.65 \pm 0.20$ \\[0.7ex]
			$r_{\rm drag} h$ [Mpc]
			& $104.64 \pm 0.81$ & $103.87 \pm 0.76$ & $102.90 \pm 0.64$ & $103.09 \pm 0.62$ & $102.97 \pm 0.61$ & $102.54 \pm 0.61$ \\[0.5ex]
			$z_\star$
			& $1114.69 \pm 0.24$ & $1114.89 \pm 0.24$ & $1110.99 \pm 0.22$ & $1110.96 \pm 0.21$ & $1110.95 \pm 0.20$ & $1111.08 \pm 0.22$ \\[0.5ex]
			$z_{\rm drag}$
			& $1083.12 \pm 0.27$ & $1083.03 \pm 0.28$ & $1078.22 \pm 0.27$ & $1078.22 \pm 0.25$ & $1078.25 \pm 0.25$ & $1078.18 \pm 0.27$ \\[0.7ex]
			$10^4\theta_\star$
			& $104.102 \pm 0.025$ & $104.092 \pm 0.024$ & $104.112 \pm 0.024$ & $104.123 \pm 0.023$ & $104.104 \pm 0.023$ & $104.108 \pm 0.024$ \\[0.5ex]
			$100Y_{\rm P}$
			& $24.605 \pm 0.0054$ & $24.602 \pm 0.0055$ & $24.590 \pm 0.0053$ & $24.591 \pm 0.0049$ & $24.591 \pm 0.0048$ & $24.589 \pm 0.0053$ \\[0.5ex]
			$\sigma_8$
			& $0.8288 \pm 0.0057$ & $0.8300 \pm 0.0057$ & $0.8218 \pm 0.0056$ & $0.8246 \pm 0.0055$ & $0.8207 \pm 0.0054$ & $0.8222 \pm 0.0056$ \\[0.5ex]
			\colrule
			\rule{0pt}{3ex}
			$\chi^2_{\rm low\text{-}\ell\ TT}$
			& 24.16        & 25.58        & 26.43    & 25.73    & 25.87    & 26.79 \\[0.5ex]
			$\chi^2_{\rm low\text{-}\ell\ EE}$
			& 395.90   & 395.95     & 396.43   & 395.69   & 395.76   & 395.68 \\[0.5ex]
			$\chi^2_{\rm high\text{-}\ell}$
			& 10539.72 & 10540.80   & 10541.83 & 10545.38 & 10545.38 & 10543.12 \\[0.5ex]
			$\chi^2_{\rm lensing}$
			& 10.12    & 8.87       & 8.81     & 8.49     & 8.41     & 9.14 \\[0.5ex]
			\colrule
			\rule{0pt}{3ex}
			$\chi^2_{\rm Planck}$
			& 10969.90 & 10971.20   & 10973.50 & 10975.29 & 10975.42 & 10974.73 \\[0.5ex]
			$\chi^2_{\rm BAO}$ 
			& $-$      & $-$        & 14.03    & 13.93    & 14.20    & 13.50 \\[0.5ex]
			$\chi^2_{\rm SN}$  
			& $-$      & 1410.41    & $-$      & $-$      & $-$      & 1408.43 \\[0.5ex]
			$\chi^2_{\rm ACT}$ 
			& $-$      & $-$        & $-$      & 244.51   & $-$      & $-$ \\[0.5ex]
			$\chi^2_{\rm SPT}$ 
			& $-$      & $-$        & $-$      & $-$      & 1885.08  & $-$ \\[0.5ex]
			\colrule
			\rule{0pt}{3ex}
			$\chi^2_{\rm total}$
			& 10969.90 & 12381.61   & 10987.53 & 11233.73 & 12874.70 & 12396.65
		\end{tabular}
	\end{ruledtabular}
	\caption{Mean values and the 68\% CL intervals of the parameters and the $\chi^2_{\rm BF}$ values for the LRA-P and LRA-PD models fit to combinations of Planck, DESI, PP, ACT, and SPT.}
	\label{t:parameters_LRA}
\end{table*}

\begin{table*}[htbp]
	\begin{ruledtabular}
		\begin{tabular}{|c|ccccc|}
			\multirow{2}{*}{\bf Parameter}
			& {\bf4-par} Planck   & {\bf4-par} Planck  & {\bf4-par} Planck   & {\bf4-par} Planck  & {\bf4-par} Planck \\
			& + PP$M_b$           & + DESI             & + DESI + PP$M_b$    & + DESI + ACT       & + DESI + SPT \\
			\colrule
			\rule{0pt}{3ex}
			$H_0$ [km/s/Mpc] 
			& $72.21 \pm 0.89$    & $69.23 \pm 0.98$   & $71.34 \pm 0.68$    & $69.90 \pm 0.91$   & $69.11 \pm 0.91$ \\[0.5ex]
			$H_0$ tension
			& $0.61\sigma$        & $2.67\sigma$       & $1.37\sigma$        & $2.27\sigma$ & $2.84\sigma$ \\[0.5ex]
			$S_8$
			& $0.799 \pm 0.010$   & $0.812 \pm 0.010$  & $0.807 \pm 0.009$   & $0.812 \pm 0.009$  & $0.811 \pm 0.010$ \\[0.5ex]
			$S_8$ tension
			& $0.44\sigma$        & $1.07\sigma$       & $0.84\sigma$        & $1.09\sigma$       & $1.02\sigma$ \\[0.5ex]
			$\Omega_{\rm m}$
			& $0.279 \pm 0.006$   & $0.298 \pm 0.007$  & $0.286 \pm 0.005$   & $0.295 \pm 0.006$  & $0.299 \pm 0.007$ \\[0.5ex]
			$\Omega_m$ tension
			& $2.85\sigma$        & $1.85\sigma$       & $2.56\sigma$        & $2.05\sigma$       & $1.81\sigma$ \\[0.5ex]
			$\Omega_{\rm m} h^2$
			& $0.146 \pm 0.001$   & $0.143 \pm 0.002$  & $0.146 \pm 0.001$   & $0.144 \pm 0.001$  & $0.143 \pm 0.001$ \\[0.5ex]
			$\tau_{\rm reio}$
			& $0.054 \pm 0.007$   & $0.055 \pm 0.007$  & $0.052 \pm 0.007$   & $0.053 \pm 0.007$  & $0.054 \pm 0.007$ \\[0.5ex]
			$n_{\rm s}$
			& $0.960 \pm 0.007$   & $0.965 \pm 0.007$  & $0.960 \pm 0.007$   & $0.962 \pm 0.007$  & $0.964 \pm 0.007$ \\[0.5ex]
			$\mathrm{log}(10^{10} A_{\rm s})$
			& $3.037 \pm 0.015$   & $3.041 \pm 0.015$  & $3.036 \pm 0.015$   & $3.042 \pm 0.014$  & $3.038 \pm 0.014$ \\[0.5ex]
			$M_b$
			& $-19.300 \pm 0.025$ & $-$                & $-19.324 \pm 0.020$ & $-$                & $-$ \\[0.5ex]
			$r_\star$ [Mpc]
			& $141.52 \pm 0.77$   & $143.75 \pm 0.93$  & $141.84 \pm 0.76$   & $143.12 \pm 0.84$  & $143.79 \pm 0.86$ \\[0.5ex]
			$r_{\rm drag}$ [Mpc]
			& $144.18 \pm 0.76$   & $146.43 \pm 0.93$  & $144.50 \pm 0.75$   & $145.80 \pm 0.84$  & $146.47 \pm 0.85$ \\[0.7ex]
			$r_{\rm drag} h$ [Mpc]
			& $104.10 \pm 0.97$   & $101.36 \pm 0.96$  & $103.08 \pm 0.69$   & $101.91 \pm 0.90$  & $101.23 \pm 0.91$ \\[0.5ex]
			$z_\star$
			& $1114.0 \pm 5.3$    & $1097.6 \pm 6.3$   & $1110.7 \pm 5.0$    & $1102.4 \pm 5.6$   & $1097.7 \pm 5.8$ \\[0.5ex]
			$z_{\rm drag}$
			& $1082.8 \pm 4.9$    & $1067.2 \pm 5.8$   & $1079.6 \pm 4.5$    & $1071.6 \pm 5.2$   & $1067.1 \pm 5.3$ \\[0.7ex]
			$10^4\theta_\star$
			& $104.131 \pm 0.034$ & $104.128 \pm 0.035$ & $104.125 \pm 0.033$ & $104.124 \pm 0.029$ & $104.110 \pm 0.030$ \\[0.5ex]
			$100Y_{\rm P}$
			& $24.601 \pm 0.007$  & $24.589 \pm 0.007$ & $24.598 \pm 0.007$  & $24.589 \pm 0.007$ & $24.587 \pm 0.007$ \\[0.5ex]
			$\sigma_8$
			& $0.828 \pm 0.008$   & $0.814 \pm 0.008$  & $0.826 \pm 0.008$   & $0.819 \pm 0.008$  & $0.812 \pm 0.007$ \\[0.5ex]
			\colrule
			\rule{0pt}{3ex}
			$A_{\rm b}$
			& $0.32 \pm 0.10$     & $< 0.219$       & $0.30^{+0.10}_{-0.12}$ & $< 0.278$          & $< 0.173$ \\[0.5ex]
			$z_{\rm b}$
			& $936 \pm 22$      & $911^{+70}_{-21}$ & $928 \pm 22$           & $925^{+34}_{-24}$  & $899^{+76}_{-20}$ \\[0.5ex]
			$\sigma_{\rm b}$
			& $153^{+30}_{-20}$   & $< 194$         & $163^{+30}_{-20}$      & $193 \pm 40$       & $194^{+40}_{-60}$ \\[0.5ex]
			$\Delta z_{\rm shift}$
			& $-43.1^{+7.6}_{-9.8}$  & $-19.5^{+13}_{-9.9}$  & $-39.1^{+7.7}_{-8.8}$  & $-27.5^{+11}_{-9.7}$ & $-18.1^{+11}_{-7.3}$ \\[0.5ex]
			\colrule
			\rule{0pt}{3ex}
			$\chi^2_{\rm low\text{-}\ell\ TT}$
			& 23.94              & 22.84               & 23.19               & 23.30              & 23.63 \\[0.5ex]
			$\chi^2_{\rm low\text{-}\ell\ EE}$
			& 395.77             & 396.42              & 395.64              & 396.23             & 395.69 \\[0.5ex]
			$\chi^2_{\rm high\text{-}\ell}$
			& 10544.56           & 10542.95            & 10545.02            & 10543.96           & 10547.73 \\[0.5ex]
			$\chi^2_{\rm lensing}$
			& 8.55               & 8.59                & 9.02                & 8.58               & 8.70 \\[0.5ex]
			\colrule
			\rule{0pt}{3ex}
			$\chi^2_{\rm Planck}$
			& 10972.82           & 10970.79            & 10972.87            & 10972.07           & 10975.74 \\[0.5ex]
			$\chi^2_{\rm BAO}$
			& $-$                & 12.91               & 14.96               & 13.09              & 13.70 \\[0.5ex]
			$\chi^2_{\rm SN}$
			& 1412.98            & $-$                 & 1410.25             & $-$                & $-$ \\[0.5ex]
			$\chi^2_{\rm SH0ES}$
			& 1.71               & $-$                 & 6.02                & $-$                & $-$ \\[0.5ex]
			$\chi^2_{\rm ACT}$
			& $-$                & $-$                 & $-$                 & $239.47$           & $-$ \\[0.5ex]
			$\chi^2_{\rm SPT}$
			& $-$                & $-$                 & $-$                 & $-$                & 1879.80 \\[0.5ex]
			\colrule
			\rule{0pt}{3ex}
			$\chi^2_{\rm total}$
			& 12387.5            & 10983.70            & 12404.11            & 11224.63           & 12869.24 \\[0.5ex]
		\end{tabular}
	\end{ruledtabular}
	\caption{Mean values and the 68\% CL intervals of the parameters and the $\chi^2_{\rm BF}$ values for the four-parameter model fit to combinations of Planck, DESI, PP$M_b$, ACT, and SPT.}
	\label{t:parameters_4model}
\end{table*}
\twocolumngrid

\FloatBarrier

\end{document}